%% file: jinst-latex-MuonIDLHCb.tex
\pdfoutput=1 
\documentclass{JINST}

\let\ifpdf\relax

\usepackage{xspace} 
\usepackage{ifthen} 
\newboolean{articletitles}
\setboolean{articletitles}{true} 

\newboolean{uprightparticles}
\setboolean{uprightparticles}{false} 
\usepackage{amssymb}
\usepackage{amsfonts}
\usepackage{upgreek} 

\input{lhcb-symbols-def} 
\input{alias} 

\title{Performance of the Muon Identification at LHCb}

\author{F.~Archilli$^a$, W.~Baldini$^b$, G.~Bencivenni$^a$, N.~Bondar$^c$, W.~Bonivento$^d$, S.~Cadeddu$^d$, P.~Campana$^a$, A.~Cardini$^d$, P.~Ciambrone$^a$, X.~Cid Vidal$^{e}$, C.~Deplano$^d$, P.~De~Simone$^a$, A.~Falabella$^{f,r}$, M.~Frosini$^{g,15}$, S.~Furcas$^{a}$\thanks{Now at Sezione INFN di Milano, Milano, Italy.}, E.~Furfaro$^h$, M.~Gandelman$^i$, J.A.~Hernando Morata$^j$, G.~Graziani$^g$, A.~Lai$^d$,  G.~Lanfranchi$^a$, J.H.~Lopes$^i$, O.~Maev$^c$, G.~Manca$^d$, G.~Martellotti$^h$, A.~Massafferri$^{k}$, D.~Milanes$^{l}$\thanks{Now at LPNHE, Universit\'e Pierre et Marie Curie, Universit\'e Paris Diderot, CNRS/IN2P3, Paris, France.}, R.~Oldeman$^{d,m}$, M.~Palutan$^a$, G.~Passaleva$^g$, D.~Pinci$^h$, E.~Polycarpo$^i$\thanks{Corresponding author.}, R.~Santacesaria$^h$, E.~Santovetti$^{n,p}$, A.~Sarti$^{a,q}$, A.~Satta$^{n}$, B.~Schmidt$^e$, B.~Sciascia$^a$, F.~Soomro$^a$, A.~Sciubba$^{h,q}$ and S. Vecchi$^{b}$\\
\llap{$^a$}INFN - Laboratori Nazionali di Frascati,\\
Frascati, Italy\\
\llap{$^b$}Sezione INFN di Ferrara,\\
Ferrara, Italy\\
\llap{$^c$}Petersburg Nuclear Physics Institute,\\
 Gatchina, St-Petersburg, Russia\\
\llap{$^d$}Sezione INFN di Cagliari,\\
Cagliari, Italy\\
\llap{$^e$}European Organisation for Nuclear Research (CERN),\\
 Geneva, Switzerland\\
\llap{$^f$}Sezione INFN di Bologna,\\
Bologna, Italy\\
\llap{$^g$}Sezione INFN di Firenze,\\
 Firenze, Italy\\
\llap{$^h$}Sezione INFN di Roma,\\
 Roma, Italy\\
\llap{$^i$}Universidade Federal do Rio de Janeiro (UFRJ),\\
Rio de Janeiro, Brasil\\
\llap{$^j$}Universidade de Santiago de Compostela,\\
 Santiago de Compostela, Spain\\
\llap{$^k$}Centro Brasileiro de Pesquisas F\'\i sicas (CBPF),\\
 Rio de Janeiro, Brasil\\
\llap{$^l$}Sezione INFN di Bari,\\
 Bari, Italy\\
\llap{$^m$}Universit\`a di Cagliari,\\
 Cagliari, Italy\\
\llap{$^n$}Sezione INFN di Roma Tor Vergata,\\
 Roma, Italy\\
\llap{$^o$}Universit\`a di Firenze,\\
 Firenze, Italy\\
\llap{$^p$}Universit\`a di Roma Tor Vergata,\\
 Roma, Italy\\
\llap{$^q$}Sapienza, Universita di Roma,\\
 Roma, Italy\\
\llap{$^r$}Universit\`a di Ferrara,\\
 Ferrara, Italy\\
E-mail: \email{poly@if.ufrj.br}}

\abstract{
The performance of the muon identification in LHCb is extracted from data using muons and hadrons produced in \Jpsimumu, \Lambdappi and $D^{\star+}\to\pi^+ D^0(K^-\pi^+)$ decays. The muon identification procedure is based on the pattern of hits in the muon chambers. A momentum dependent binary requirement is used to reduce the probability of hadrons to be misidentified as muons to the level of 1\%, keeping the muon efficiency in the range of 95-98\%. As further refinement, a likelihood is built for the muon and non-muon hypotheses. Adding a requirement on this likelihood that provides a total muon efficiency at the level of 93\%, the hadron \misids are below 0.6\%.}


\begin{document}

\input{introduction}

\input{principles}

\input{method}

\input{results}

\input{conclusion}

\acknowledgments

We express our gratitude to our colleagues in the CERN accelerator
departments for the excellent performance of the LHC. We thank the
technical and administrative staff at CERN and at the LHCb institutes,
and acknowledge support from the National Agencies: CAPES, CNPq,
FAPERJ and FINEP (Brazil); CERN; NSFC (China); CNRS/IN2P3 (France);
BMBF, DFG, HGF and MPG (Germany); SFI (Ireland); INFN (Italy); FOM and
NWO (The Netherlands); SCSR (Poland); ANCS (Romania); MinES of Russia and
Rosatom (Russia); MICINN, XuntaGal and GENCAT (Spain); SNSF and SER
(Switzerland); NAS Ukraine (Ukraine); STFC (United Kingdom); NSF
(USA). We also acknowledge the support received from the ERC under FP7
and the Region Auvergne.

\end{document}

%% file: lhcb-symbols-def.tex
\def\lhcb {LHCb\xspace}
\def\ux85 {UX85\xspace}

\ifthenelse{\boolean{uprightparticles}}%
{

 \def\Ppi         {\ensuremath{\uppi}\xspace}

 \def\Ppsi        {\ensuremath{\uppsi}\xspace}

 \def\PDelta      {\ensuremath{\Delta}\xspace}                 
 \def\PXi      {\ensuremath{\Xi}\xspace}                 
 \def\PLambda      {\ensuremath{\Lambda}\xspace}                 
 \def\PSigma      {\ensuremath{\Sigma}\xspace}                 
 \def\POmega      {\ensuremath{\Omega}\xspace}                 
 \def\PUpsilon      {\ensuremath{\Upsilon}\xspace}                 
 
 \def\PB      {\ensuremath{\mathrm{B}}\xspace}                 
                  
 \def\PD      {\ensuremath{\mathrm{D}}\xspace}

 \def\PJ      {\ensuremath{\mathrm{J}}\xspace}                 
 \def\PK      {\ensuremath{\mathrm{K}}\xspace}

 \def\Pi      {\ensuremath{\mathrm{i}}\xspace}

}
{

 \def\Ppi         {\ensuremath{\pi}\xspace}

 \def\Ppsi        {\ensuremath{\psi}\xspace}                 
                  
 \mathchardef\PDelta="7101
 \mathchardef\PXi="7104
 \mathchardef\PLambda="7103
 \mathchardef\PSigma="7106
 \mathchardef\POmega="710A
 \mathchardef\PUpsilon="7107
                  
 \def\PB      {\ensuremath{B}\xspace}                 
                  
 \def\PD      {\ensuremath{D}\xspace}

 \def\PJ      {\ensuremath{J}\xspace}                 
 \def\PK      {\ensuremath{K}\xspace}

 \def\Pi      {\ensuremath{i}\xspace}

}

\def\pion  {\ensuremath{\Ppi}\xspace}

\def\pip   {\ensuremath{\pion^+}\xspace}

\def\kaon  {\ensuremath{\PK}\xspace}
  \def\Kbar  {\kern 0.2em\overline{\kern -0.2em \PK}{}\xspace}

\def\Kz    {\ensuremath{\kaon^0}\xspace}
\def\Kzb   {\ensuremath{\Kbar^0}\xspace}
\def\KzKzb {\ensuremath{\Kz \kern -0.16em \Kzb}\xspace}
\def\Kp    {\ensuremath{\kaon^+}\xspace}
\def\Km    {\ensuremath{\kaon^-}\xspace}

\def\KpKm  {\ensuremath{\Kp \kern -0.16em \Km}\xspace}

  \def\Dbar    {\kern 0.2em\overline{\kern -0.2em \PD}{}\xspace}
\def\D       {\ensuremath{\PD}\xspace}

\def\Dz      {\ensuremath{\D^0}\xspace}
\def\Dzb     {\ensuremath{\Dbar^0}\xspace}
\def\DzDzb   {\ensuremath{\Dz {\kern -0.16em \Dzb}}\xspace}
\def\Dp      {\ensuremath{\D^+}\xspace}
\def\Dm      {\ensuremath{\D^-}\xspace}

\def\DpDm    {\ensuremath{\Dp {\kern -0.16em \Dm}}\xspace}

\def\Dstarp  {\ensuremath{\D^{*+}}\xspace}

  \def\Bbar    {\kern 0.18em\overline{\kern -0.18em \PB}{}\xspace}

\def\jpsi     {\ensuremath{{\PJ\mskip -3mu/\mskip -2mu\Ppsi\mskip 2mu}}\xspace}

  \def\Y#1S{\ensuremath{\PUpsilon{(#1S)}}\xspace}

\newcommand{\decay}[2]{\ensuremath{#1\!\to #2}\xspace}         

\def\to                 {\ensuremath{\rightarrow}\xspace}

\def\AT#1     {\ensuremath{A_{\mathrm{T}}^{#1}}\xspace}           

\def\C#1      {\ensuremath{\mathcal{C}_{#1}}\xspace}                       
\def\Cp#1     {\ensuremath{\mathcal{C}_{#1}^{'}}\xspace}                    
\def\Ceff#1   {\ensuremath{\mathcal{C}_{#1}^{\mathrm{(eff)}}}\xspace}        
\def\Cpeff#1  {\ensuremath{\mathcal{C}_{#1}^{'\mathrm{(eff)}}}\xspace}       
\def\Ope#1    {\ensuremath{\mathcal{O}_{#1}}\xspace}                       
\def\Opep#1   {\ensuremath{\mathcal{O}_{#1}^{'}}\xspace}                    

\newcommand{\tev}{\ensuremath{\mathrm{\,Te\kern -0.1em V}}\xspace}
\newcommand{\gev}{\ensuremath{\mathrm{\,Ge\kern -0.1em V}}\xspace}
\newcommand{\mev}{\ensuremath{\mathrm{\,Me\kern -0.1em V}}\xspace}
\newcommand{\kev}{\ensuremath{\mathrm{\,ke\kern -0.1em V}}\xspace}
\newcommand{\ev}{\ensuremath{\mathrm{\,e\kern -0.1em V}}\xspace}
\newcommand{\gevc}{\ensuremath{{\mathrm{\,Ge\kern -0.1em V\!/}c}}\xspace}
\newcommand{\mevc}{\ensuremath{{\mathrm{\,Me\kern -0.1em V\!/}c}}\xspace}
\newcommand{\gevcc}{\ensuremath{{\mathrm{\,Ge\kern -0.1em V\!/}c^2}}\xspace}
\newcommand{\gevgevcccc}{\ensuremath{{\mathrm{\,Ge\kern -0.1em V^2\!/}c^4}}\xspace}
\newcommand{\mevcc}{\ensuremath{{\mathrm{\,Me\kern -0.1em V\!/}c^2}}\xspace}

\def\invfb   {\ensuremath{\mbox{\,fb}^{-1}}\xspace}

\def\gsim{{~\raise.15em\hbox{$>$}\kern-.85em
          \lower.35em\hbox{$\sim$}~}\xspace}
\def\lsim{{~\raise.15em\hbox{$<$}\kern-.85em
          \lower.35em\hbox{$\sim$}~}\xspace}

\def\pt         {\mbox{$p_{\rm T}$}\xspace}

\def\gauss      {\mbox{\textsc{Gauss}}\xspace}

\def\tell1  {TELL1\xspace}
\def\ukl1   {UKL1\xspace}



%% file: alias.tex
\newcommand{\particle}[1]{{\ensuremath{\rm #1}}}

\newcommand{\Jpsi}{\particle{J\!/\!\psi}\,\,}
\newcommand{\Lambdappi}{\ensuremath{\Lambda^0\rightarrow p\pi^- }\,\,}

\newcommand{\Jpsimumu}{\ensuremath{J/\psi\rightarrow \mu^+\mu^- }\,\,}
\newcommand{\mprobe}{\ensuremath{\,\mu_{\rm probe}}\,\,}

\newcommand{\misid}{\ensuremath{{\rm misidentification}\,\,{\rm probability}}\xspace}
\newcommand{\misids}{\ensuremath{{\rm misidentification}\,\,{\rm probabilities}}\xspace}
\newcommand{\Dkpi}{\decay{\Dz}{K^- \pi^+}}

\newcommand{\eIM}{\ensuremath{\varepsilon_{IM}\,\,}}
\newcommand{\ensh}{\ensuremath{\varepsilon_{\rm NShared}\,\,}}
\newcommand{\RIM}{\ensuremath{\wp_{IM}\,\,}}
\newcommand{\RIMpi}{\ensuremath{\wp_{IM}(\pi\rightarrow\mu)\,\,}}
\newcommand{\RIMka}{\ensuremath{\wp_{IM}(K\rightarrow\mu)\,\,}}
\newcommand{\RIMpr}{\ensuremath{\wp_{IM}(p\rightarrow\mu)\,\,}}
\newcommand{\eDLL}{\ensuremath{\varepsilon_{\rm muDLL}\,\,}}
\newcommand{\eCDLL}{\ensuremath{\varepsilon_{\rm DLL}\,\,}}

\newcommand{\RDLL}{\ensuremath{\wp_{\rm muDLL}\,\,}}

\newcommand{\figref}[1]{Fig.~\ref{#1}}
\newcommand{\tabref}[1]{Table~\ref{#1}}
\newcommand{\secref}[1]{Section~\ref{#1}}

%% file: introduction.tex
\section{Introduction}
\label{sec:intro}

LHCb~\cite{Alves:2008zz} is a dedicated heavy flavour experiment, designed to exploit the high $pp\to c\bar{c}$ and $pp\to b\bar{b}$ cross-sections at the LHC in order to perform precision measurements of CP violation and rare decays. Muons are present in the final state of many of the key decays, sensitive to new physics, as shown, for example, in
~\cite{bsmumu,physics2,physics3,physics5,physics6}, among others. 
Moreover, they play a crucial role in the determination of the flavor tagging of the neutral $B$ mesons and are also present in the signatures of interesting electroweak and strong processes. The muon identification procedure must provide high muon efficiency while keeping the incorrect identification probability of hadrons as muons (misidentification probabilities) at the lowest possible level. The pion misidentification is one of the major sources of combinatoric background for decays with muons in the final state. It is also important to keep the other hadron misidentification probabilities at low levels 
 so that rare decays can be separated from more abundant hadronic decays 
with similar or identical topology. 

       This paper presents the performance of the muon identification in LHCb, obtained from the data recorded in 2011, corresponding to approximately 1\invfb. In \secref{sec:detector}, a brief description of the LHCb spectrometer and the muon detection system is given. The muon identification algorithm is discussed in \secref{sec:principles}. The method used to extract the muon efficiency and the misidentification probability from data is explained in \secref{sec:method}. Finally, the performance results are presented in \secref{sec:results}, followed by the conclusions in \secref{sec:conclusions}. 
       
\section{The LHCb experiment and the muon system}
\label{sec:detector}

The LHCb detector~\cite{Alves:2008zz} is a single-arm forward spectrometer. A vertex locator (VELO) determines with high precision the positions of the vertices of $pp$ collisions (PVs) and the decay vertices of long-lived particles. The tracking system includes a silicon strip detector located in front of a dipole magnet with an integrated field of about 4~Tm, and a combination of silicon strip detectors and straw drift chambers placed behind the magnet. The momentum of charged particles is determined with a resolution of $\sigma_p/p\sim$0.4(0.6)\% at a momentum scale of 3(100)\gevc. 

Charged hadron identification is achieved with two ring-imaging Cherenkov (RICH) detectors. The calorimeter system consists of a scintillator pad detector, a preshower, an electromagnetic calorimeter and a hadronic calorimeter. It identifies high transverse energy\footnote{The transverse energy of a 2$\times$2 cells cluster is defined as 
$E_T =\displaystyle\sum_{i=1}^4 E_i \sin{\theta_i}$, 
where $E_i$
is the energy deposited in cell $i$ and $\theta_i$ is the angle between the z-axis and a neutral particle
assumed to be coming from the mean position of the interaction envelope hitting the centre of the cell~\cite{triggerPaper}.} hadron, electron and photon candidates and provides information for the trigger. 

\begin{figure}
\centering
\includegraphics[width=0.7\textwidth]{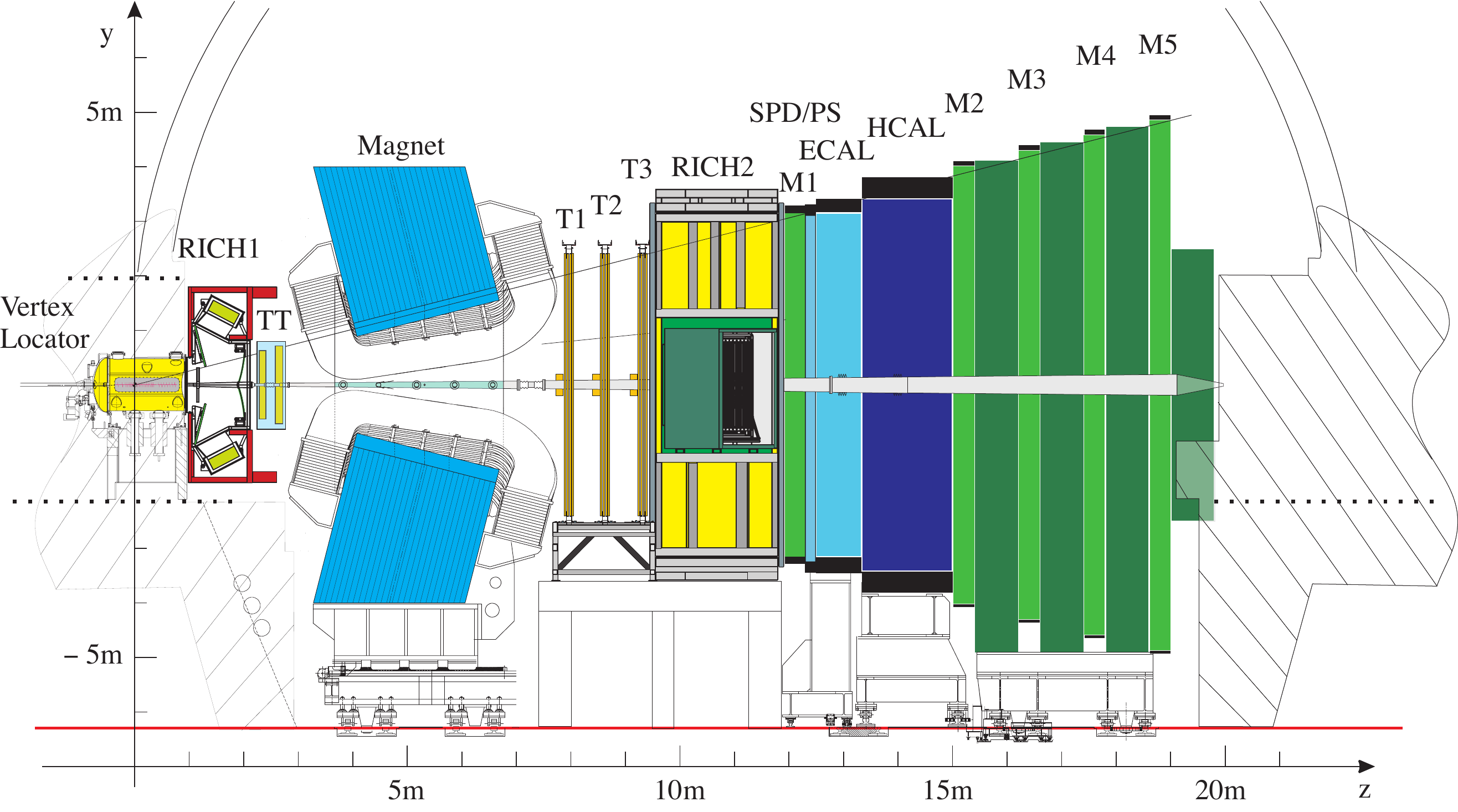}
\caption{Schematic view of the LHCb experiment. The muon stations are seen as the five green vertical bars, the second one placed just after the calorimeters, shown as the blue rectangles.}
\protect\label{fig:lhcbdet}
\end{figure}

The muon system~\cite{ref:muonSystem} is composed of five stations (M1-M5) of rectangular shape, placed along the beam axis, as shown in ~\figref{fig:lhcbdet}. Station M1 is located in front of the calorimeters and is used to improve the transverse momentum measurement in the first level hardware trigger. Stations M2 to M5 are placed downstream the calorimeters and are interleaved with iron absorbers 80 cm thick to select penetrating muons. 
The total absorber thickness in front of station M2, including the calorimeters, is approximately 6.6 interaction lengths. 
More than 99\% of the total area of the system is equipped with multi-wire proportional chambers (MWPC) with Ar/CO2/CF4(40:55:5) as gas mixture. Only the inner part of the first station is instrumented with triple-GEM detectors filled with Ar/CO2/CF4(45:15:40). 

\begin{figure}
\centering
\includegraphics[width=0.5\textwidth]{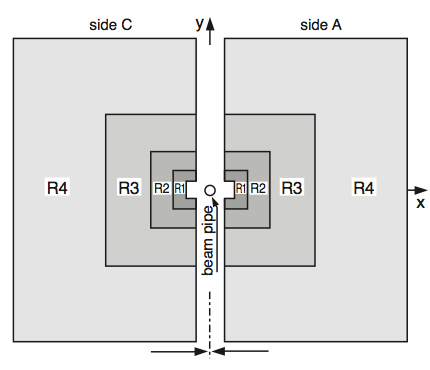}
\caption{Schematic view of one muon system station (reproduced from~\cite{ref:muonSystem}).} 
\label{fig:msregions}
\end{figure}

 The chambers are positioned to provide with their sensitive area a hermetic geometric acceptance to high momentum particles coming from the interaction point. In addition, the chambers of different stations form projective towers pointing to the interaction point.  The detectors provide digital space point measurements on the particle trajectories, supplying information to the trigger processor and to the data acquisition (DAQ). The information is obtained by partitioning the detector into rectangular logical pads whose dimensions define the x, y resolution in the plane perpendicular to the beam axis. Each station is divided into four regions, R1 to R4 with increasing distance from the beam axis, as shown in~\figref{fig:msregions}. The linear dimensions of the regions R1, R2, R3, R4, and their segmentation scale in the ratio 1:2:4:8. 

 Each muon station is designed to perform with an efficiency above 99\% in a 20$\,$ns time window with a noise rate below 1$\,$kHz per physical channel, which was achieved during operation, as described in~\cite{ref:muonSystem}.

The muon system provides information for the selection of high transverse momentum muons at the trigger level and for the offline muon identification. This document  refers to the latter procedure, which uses only the information from the 4 stations located after the calorimeters. The muon identification in the trigger system is described in~\cite{LHCb-PUB-2011-017}.

%% file: principles.tex
\section{The muon identification procedure}
\label{sec:principles}

The muon identification strategy can be divided in three steps: 
\begin{itemize}
\item A loose binary selection of muon candidates based on the penetration of the muons through the calorimeters and iron filters, which provides high efficiency while reducing the misidentification probability of hadrons to the percent level (called IsMuon); 
\item Computation of a likelihood for the muon and non-muon hypotheses, based on the pattern of hits around the extrapolation to the different muon stations of the charged particles trajectories reconstructed with high precision in the tracking system. The logarithm of the ratio between the muon and non-muon hypotheses is used as discriminating variable and called muDLL.  
\item Computation of a combined likelihood for the different particle hypotheses, including information from the calorimeter and RICH systems.  The logarithm of the ratio between the muon and pion hypotheses is used as discriminating variable and called DLL. 
\end{itemize}
Additionally the number of tracks identified as muons that share a hit with a given muon candidate (called NShared) can be used to further reject false candidates.

\subsection{IsMuon binary selection}

The binary selection is defined according to the number of stations where a hit is found within a field of interest (FOI) defined around the track extrapolation. The number of stations required to have a muon signal is a function of track momentum ($p$), as shown in~\tabref{tab:ismuonDef}. The sizes of the fields of interest also depend on the particle momentum and are defined according to the expected multiple scattering suffered by a muon when traversing the material. The FOI are parameterized separately for the 4 regions of the 4 different stations downstream the calorimeter in both $x$ and $y$ directions according to: 
\begin{equation}
 \mathrm{FOI}= a + b\times \exp(-c\times p).
\end{equation}
The parameters $a$, $b$ and $c$ have been determined using muons from a full detector Monte Carlo simulation~\cite{LHCb-PROC-2011-006}.  

\begin{table}[htb]
\begin{center}
\begin{tabular}{|l|c|}\hline
 Momentum range       & Muon stations \\\hline\hline
 3 \gevc $<  p <$ 6 \gevc  & M2 and M3  \\\hline
 6 \gevc $<  p <$ 10 \gevc & M2 and M3 and (M4 or M5) \\\hline
 $ p >$ 10 \gevc & M2 and M3 and M4 and M5 \\\hline
\end{tabular}\end{center}
\caption{Muon stations required to trigger the
IsMuon decision as a function of momentum range.}
\label{tab:ismuonDef}
\end{table}

For tracks passing the IsMuon requirement, the muon identification can be further improved by a selection based on the logarithm of the ratio between the likelihoods for the muon and non-muon hypotheses (muDLL). 

\subsection{Muon and non-muon likelihoods}
\label{sec:muonll}
The likelihoods are computed as the cumulative probability distributions of 
 the average squared distance significance $D^2$ of the hits in the muon chambers 
with respect to the linear extrapolation of the tracks from the tracking system. True muons tend to have a much narrower $D^2$ distribution, close to zero, than the other particles that are incorrectly selected by the IsMuon requirement. 

The average squared distance significance is defined as:

\begin{equation}
  D^2 = {1 \over N} \sum_{i}  \left\{
 \left( {x^i_{closest} - x^i_{track} \over pad^i_x} \right )^2 + 
\left({y^i_{closest} - y^i_{track} \over pad^i_y} \right) ^2 \right \}
\label{eq:distance_new}
\end{equation}

\noindent where the index $i$ runs over the stations containing hits within the FOI,
 ($x^i_{closest,i}, y^i_{closest}$) are the coordinates
 of the closest hit to the track extrapolation point for each station ($x^i_{track},y^i_{track}$) and $pad^i_{x,y}$ correspond to one half of the pad sizes in the x,y directions. The total number of stations containing hits within their FOI is denoted by $N$. 

The $D^2$ distribution for muons depends on the multiple scattering and, therefore, on the momentum ($p$) and polar angle ($\theta$) distributions of the analyzed sample. In order to avoid a dependence of the muon likelihood on the calibration sample (with particular $p$ and $\theta$), the tuning of the muon likelihood is performed separately in momentum bins and muon detector regions (which correspond to 4 intervals in $\theta$). 

The likelihood for the non-muon hypothesis is calibrated with the $D^2$ distribution for protons, since the other charged hadrons (pions or kaons) selected by IsMuon will present a $D^2$ distribution with a component identical to the protons and a component very similar to the true muons, due to decays in flight before the calorimeter. 
For protons, the hits in the muon system found around the track extrapolation are essentially due to three sources: hits from punch-though~\cite{detector} protons, hits from true muons pointing to the same direction of the proton or random hits. The last two are at first order uncorrelated to the proton momentum while the first one can present some momentum dependence, less important however than the dependence expected for muons. 

Hence, the tuning of the non-muon likelihood is merely performed separately for the 4 muon system regions, due to their different granularity. 

The likelihood for the muon (or non-muon) hypothesis is then defined, for each candidate, as the integral of the calibrated muon (or proton) $D^2$ probability density function from 0 to the measured value, $D^2_0$. 

The results presented in this document are obtained with a muon likelihood calibrated with muons from \Jpsimumu decays selected from the data taken in 2010,  as described in~\secref{sec:method}. The non-muon likelihood has been calibrated with a simulated sample of decays \Lambdappi.

\begin{figure}[!hp]
\centering\begin{tabular}{cc}
\includegraphics[width=.46\textwidth]{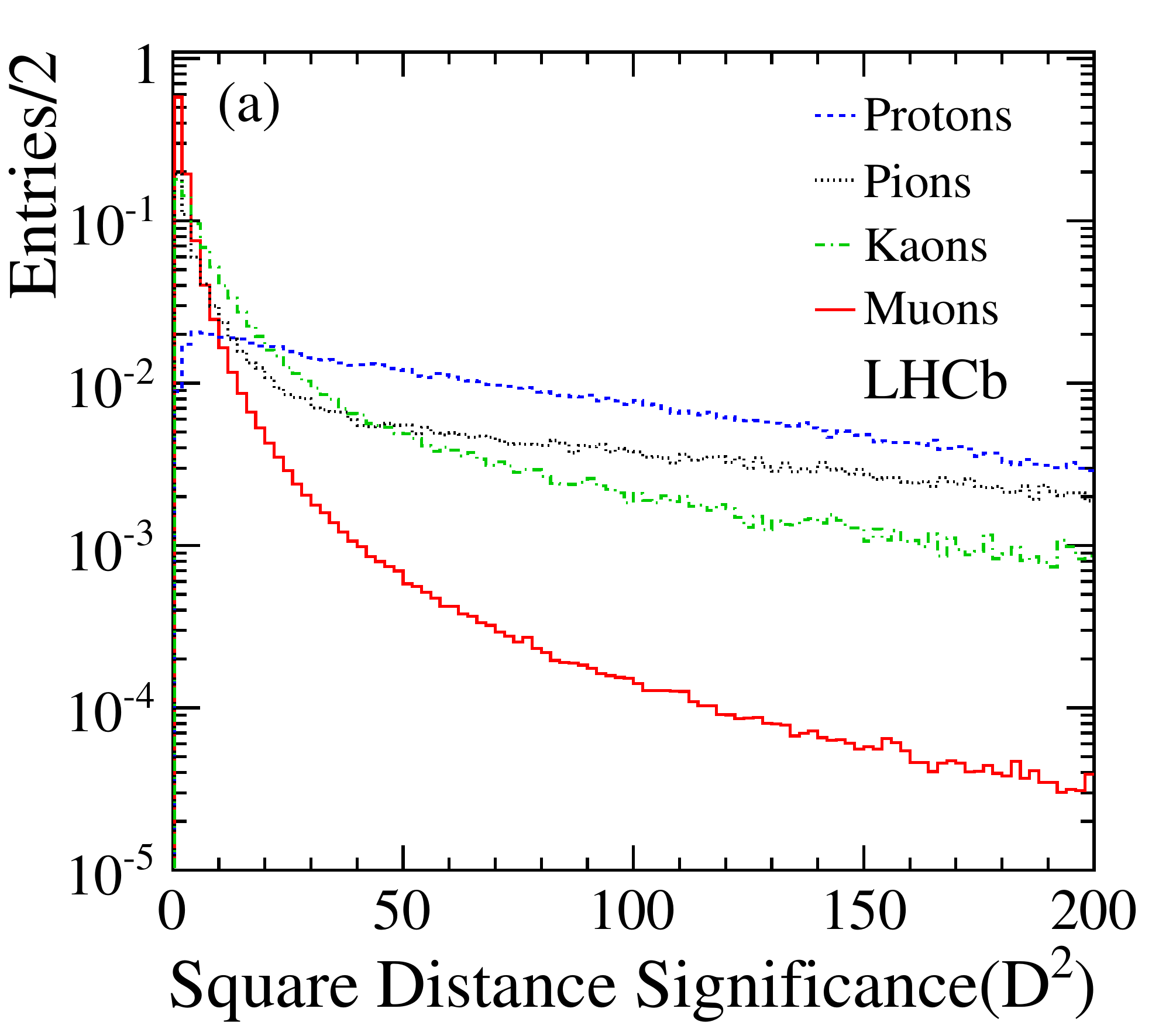} 
\includegraphics[width=.46\textwidth]{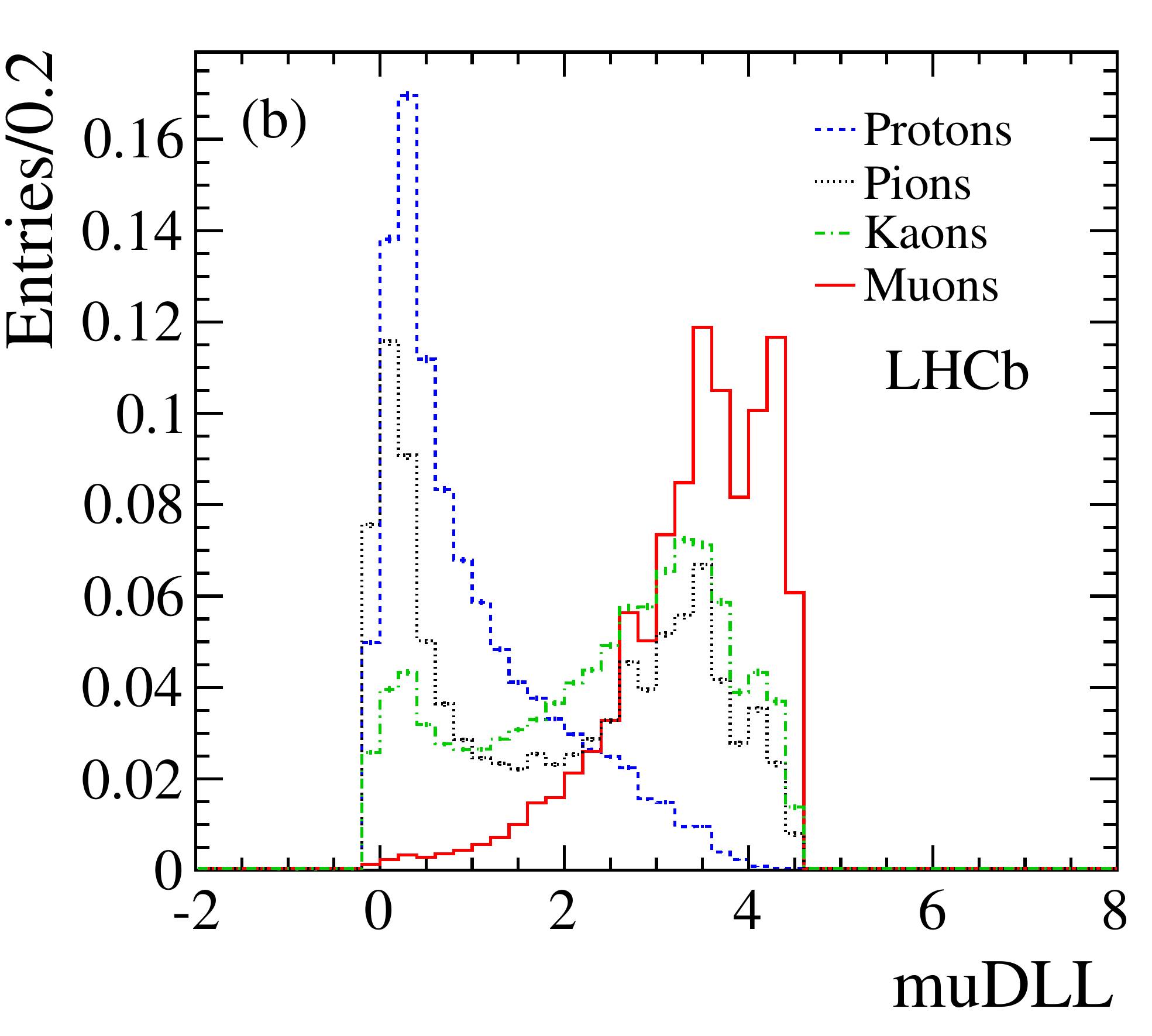}\\
\end{tabular}
\vspace{-0.5cm}
\caption{Average square distance significance distributions for muons, protons, pions and kaons (a) and the corresponding muDLL distributions (b).}
\label{fig:mudist&dll}
\end{figure}

The $D^2$ distributions for muons, protons, pions and kaons obtained from data are shown in \figref{fig:mudist&dll}(a). The distributions of the logarithm of the ratio between the muon and non-muon hypotheses (muDLL) are shown in \figref{fig:mudist&dll}(b). More details about the selection of the particles used to make these plots and to extract the performance are given in~\secref{sec:method}.  

\subsection{Combined likelihoods}

The muon and non-muon likelihoods presented in~\secref{sec:muonll} can be combined with the likelihoods provided by the RICH systems and the calorimeters to improve the muon identification performance. 

The Cherenkov angles measured in the two RICH detectors are combined with the track momentum using an overall event log-likelihood algorithm. For each track in the event, a likelihood is assigned to each of the different mass hypotheses (electron, muon, pion, kaon and proton). The RICH likelihood can differentiate between muon and other particles in particular at low momentum, below 5$\,$\gevc~\cite{richPaper}.
 
The energy deposition in the calorimeters also allows the evaluation of likelihoods for the muon  (minimum ionizing particle), electron and hadron hypotheses.

A combined log-likelihood is then obtained for each track and for each of the different mass hypotheses by summing the logarithms of the likelihoods obtained using the muon system, the RICH and the calorimeters. In this computation, the non-muon likelihood obtained in the muon system is assigned to the electron, pion, kaon and proton hypotheses. The difference of the combined log-likelihoods for the muon and pion hypotheses (DLL) is then used to identify the muons.  

\subsection{Discriminating variable based on hits sharing}
\begin{figure}[!hp]
\centering
\includegraphics[width=.46\textwidth]{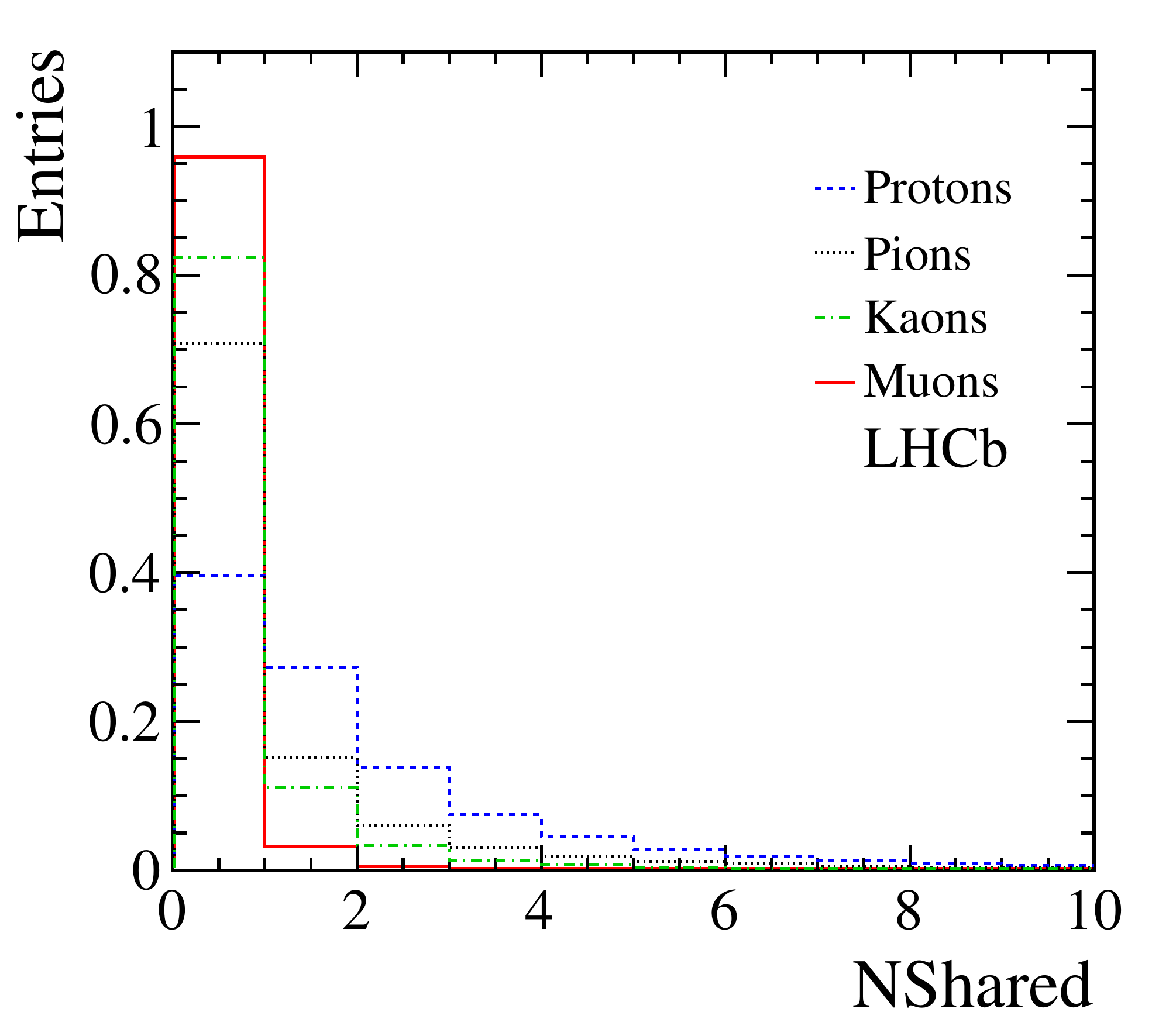}
\vspace{-0.5cm}
\caption{Normalized NShared distributions for muons, protons, kaons and pions.}
\label{fig:nshDist}
\end{figure}

Different tracks can be associated to the same muon hits when the matching of tracks to muon chamber hits is performed. Reducing the number of tracks that share hits can help to improve the misidentification probability.
To use this information, a discriminant variable named NShared is built for tracks satisfying the IsMuon criteria
and a score of 1 is added to a given track if it shares any hits with another one. The score is given to the track to which the
hit is more distant. With this definition, a track having NShared=3, for example, shares at least one hit with 3 other tracks in the event, all of them with $D^2$ values smaller than the track own $D^2$.  
Selecting muons with NShared=0 is the usual way to reduce the probability of incorrectly identifying hadrons as muons due to nearby true muons in high multiplicity events, but looser requirements can also be applied as shown in 
\figref{fig:nshDist}. 

%% file: method.tex
\section{Method for the extraction of efficiencies}
\label{sec:method}

In order to extract the performance of the muon identification from data, muon, proton, pion, and kaon candidates are selected with high purity from two body decays using kinematical requirements only. When necessary, the purity is improved by using a {\it tag and probe technique} where particle identification requirements are applied to one of the tracks (tag) while the other (probe) is used for the computation of the muon efficiency or of the hadron \misid.  

\subsection{Selection of control samples}

An abundant source of muons is provided in the experiment by the \Jpsimumu decay. By requiring the muons to have a high impact parameter with respect to the primary vertex and the reconstructed \jpsi to have a large flight distance significance and good decay vertex quality, most of the combinatorial background originating from the tracks coming from the primary vertex is removed and the sample gets enriched by $B\to J/\psi X$ candidates. In order to reduce further the combinatorial background, one of the muons is required to be identified as a muon. This is defined as the {\it tag} muon, while the one being probed is only required to have \pt$>0.8\,$\gevc. 

Protons are selected from the \Lambdappi decays reconstructed using decay vertex quality criteria and detachment of the decay vertex from the primary one. Besides, the invariant mass obtained by assigning the $\pi$ mass to the two daughters is required to be out of a window of 20 \mevcc around the nominal $K^0_s$ mass. 

The \Dstarp$\to$\pip\Dz($\rightarrow$\Km\pip) decays are the source of pions and kaons. Once again relatively high impact parameter is required for the daughters of the \Dz while the \Dz flight direction is required to point to the primary vertex. To evaluate the pion misidentification probability, the tag kaon is selected  using a suitable cut on the \pion-\kaon log-likelihoods difference, based on the RICH information. To evaluate the kaon misidentification probability, the RICH particle identification is used to identify the pion. Quality criteria are used for the \Dstarp and \Dz decay vertices. A window of $25\,$\mevcc around the nominal \Dz mass is used to exclude the doubly Cabibbo suppressed mode and the $K^+K^-$ and $\pi^+\pi^-$ decay channels. 

To avoid potential biases from the trigger requirements, in the \jpsi and $\Lambda^0$ samples only events triggered independently on the probe track are used; this condition has to be satisfied at both hardware and software level, as explained in ~\cite{triggerPaper}. For the \Dkpi sample, a substantial fraction of the events would be lost by such requirement. Therefore the hardware trigger is required to be activated independently on the probe track (kaon or pion) and the software trigger decision is based on impact parameter and detachment from the primary vertex only, with no particle identification requirement. 

After the background subtraction of selected two-body decays, the number of muon, proton, pion and kaon candidates in the 2011 data samples are 2.4, 16.1, 11.7 and 12.3 millions, respectively.

\subsection{Efficiency evaluation}

As a baseline method to evaluate the efficiency $\epsilon_{muonID}$ of a generic muon identification requirement denoted in this section by $muonID$ (e.g. IsMuon true or DLL greater than a given cut), is used :
\begin{equation}\epsilon_{muonID}=\frac{S_{true}}{S_{true}+S_{false}},
\label{eq:Eff1}
\end{equation}

where $S_{true}$ and $S_{false}$ are the numbers of signal events satisfying and not satisfying $muonID$, extracted from data using
\begin{equation}S_{true,false}= N_{true,false}-B_{true,false}.
\end{equation}

$N_{true,false}$ are obtained by counting the number of \jpsi candidates with invariant mass lying within a signal mass window around the \jpsi mass; the number of background events within the same mass window, $B_{true,false}$, is computed by extrapolating to the signal window the mass fit done in the \jpsi sidebands.

For the proton \misid, the same method is used. The kaon and pion \misids are also obtained with Eq.~\ref{eq:Eff1}, but $S_{true,false}$ and $B_{true,false}$ are extracted directly from a full fit of the signal and background shapes to the invariant mass distribution of the $D^{0}$ candidates.

%% file: results.tex
\section{Results}
\label{sec:results}

The muon identification performance is presented in terms of the muon efficiency and hadron \misids for the different requirements. In all cases, the performance is evaluated for tracks extrapolated within the geometrical acceptance of the muon detector.

\subsection{Performance of the IsMuon binary selection}

\begin{figure}[hb!]
\begin{tabular}{cc}
\includegraphics[angle=0,width=0.48\textwidth]{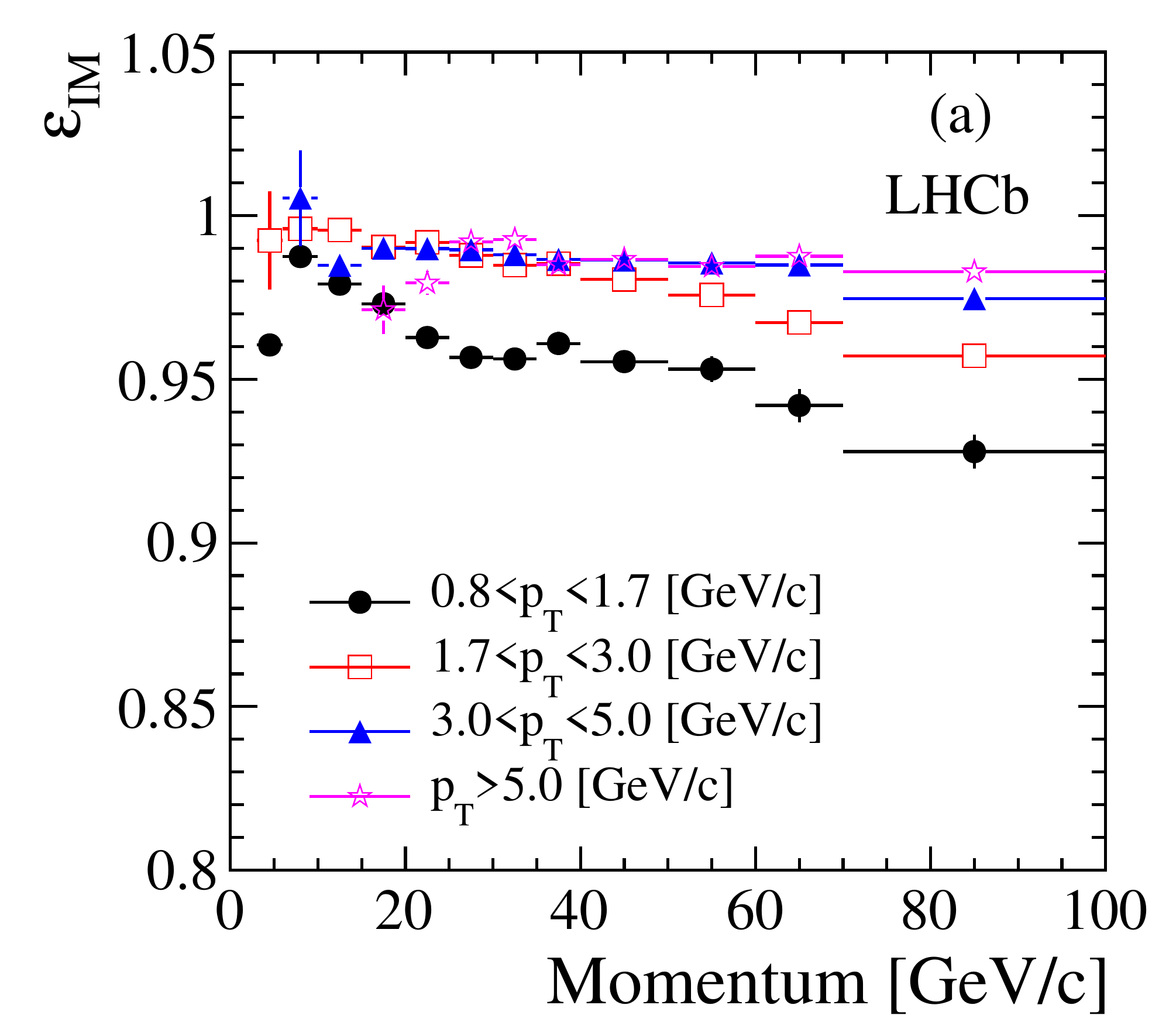}&
\includegraphics[angle=0,width=0.48\textwidth]{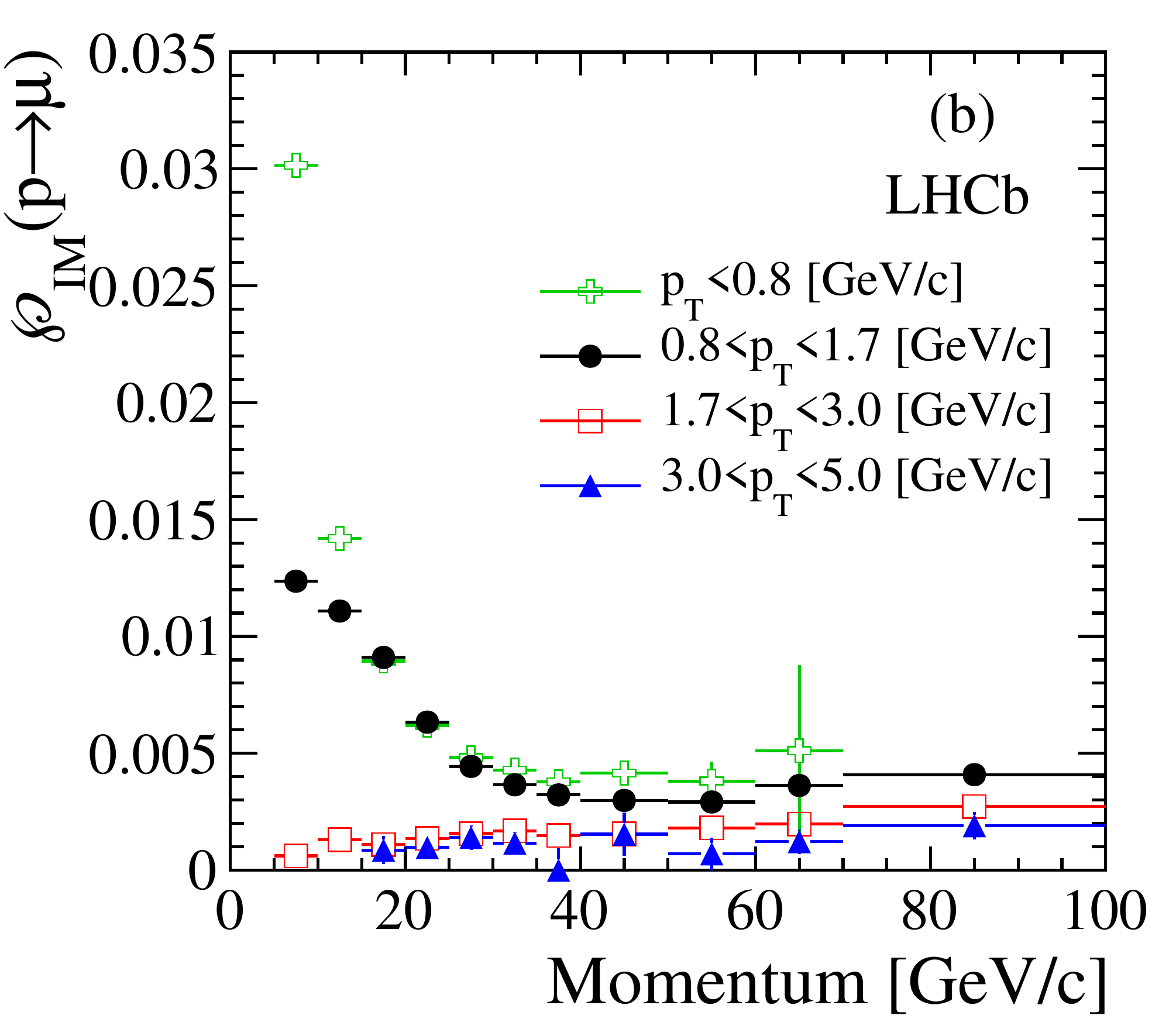}\\
\includegraphics[angle=0,width=0.48\textwidth]{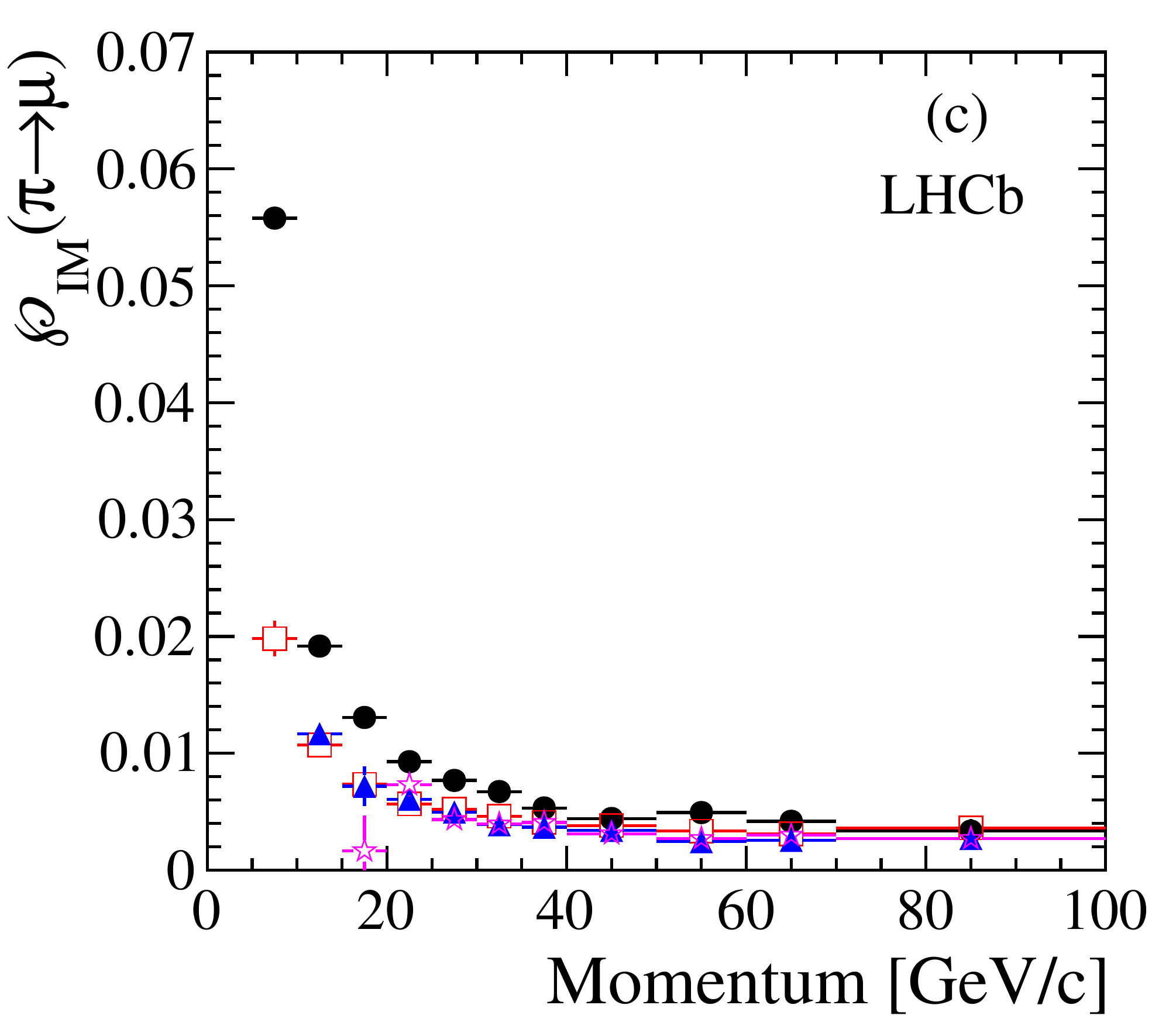}&
\includegraphics[angle=0,width=0.48\textwidth]{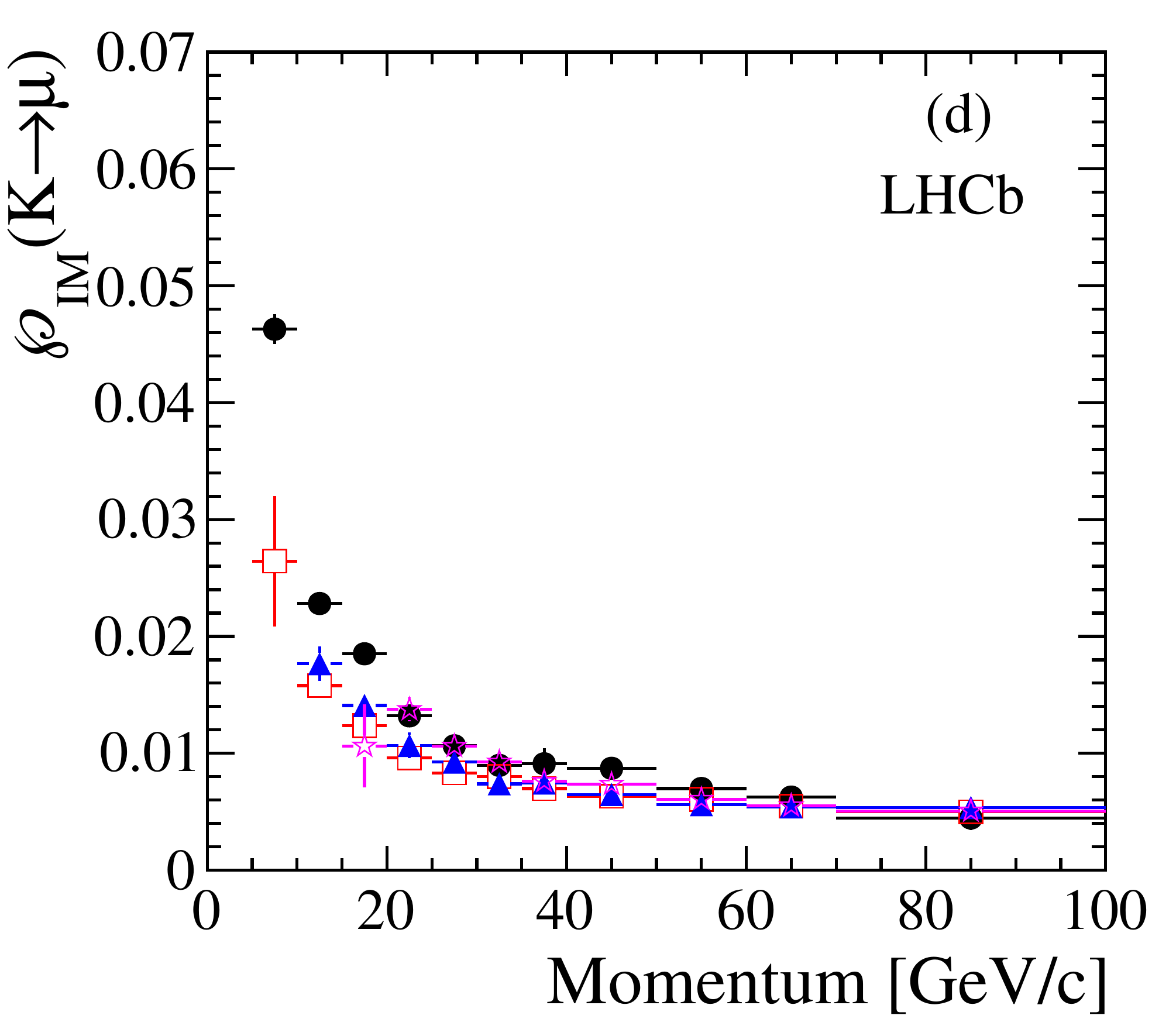}\\
\end{tabular}
  \caption{IsMuon efficiency and misidentification probabilities, as a function of momentum, in ranges of transverse momentum: \eIM (a), \RIMpr (b), \RIMpi (c) and \RIMka (d).}
  \label{fig:IMperfVSppt}
\end{figure}

The efficiency of the IsMuon requirement, \eIM, is the efficiency of finding hits within the fields of interest in the muon chambers for tracks extrapolated to the muon system. In \figref{fig:IMperfVSppt}, \eIM is shown as a function of the muon momentum, for different transverse momentum ranges. A weak dependency with transverse momentum is observed and in particular a drop of $\sim$2\% is measured for the lowest \pt interval. This efficiency drop is essentially due to tracks close to the inner edges of region R1 which in principle have their extrapolation points within M1 and M5 acceptance, but are in fact scattered outside the detector. For particles with \pt above 1.7$\,$\gevc, the efficiency is above 97\% in the whole momentum range, from 3$\,$\gevc to 100$\,$\gevc. The average efficiency obtained for the \mprobe in the \Jpsi calibration sample is \eIM=($98.13 \pm 0.04)$\%, for  particles with $p>3\,$\gev and \pt$>0.8\,$\gevc. 

The \misids \RIMpr, \RIMpi and \RIMka are also shown in \figref{fig:IMperfVSppt}. The observed decrease of \RIM  with increasing transverse momentum is expected, since tracks with higher transverse momentum traverse the detector at higher polar angles, in the lower occupancy regions. The proton \misid is smaller than 0.5\% for all $p_T$ ranges and momentum above 30\gevc. It drops quickly with momentum for the lowest \pt ranges, reaching a plateau at about 30-40$\,$\gevc. 
The pion and kaon \misids have a similar behavior, increasing with decreasing $p_T$. Above 40$\,$\gevc, the pion \misid is almost at the level of the proton \misid. At low momentum, decays in flight are the dominant source of incorrect identification, as can be seen from the difference between the pion/kaon and proton curves. While the proton \misid, within the \pt intervals chosen, lies within 0.1-1.3\%, the pion and kaon \misids are within 0.2-5.6\% and 0.6-4.5\%, respectively. For momentum above 30$\,$\gevc, \RIMpi and \RIMka have a small dependence on \pt. At the lowest $p_T$ range, the kaon \misid is lower than the pion for the lowest momentum interval, in spite of the larger decay width of kaons to muons. Since the muon is produced with a larger opening angle with respect to the original track trajectory in kaon decays than in pion decays, and on average low momentum particles tend to decay further upstream in the detector, then the hits in the muon chambers have a higher probability to lie outside the fields of interest.

When integrated over $p>3\,$\gevc and the whole \pt spectra of our calibration samples, the average values for the \misids are \RIMpr=(1.033 $\pm$ 0.003)\%, \RIMpi=(1.025$\pm$0.003)\% and \RIMka=(1.111$\pm$0.003)\%. For pions and kaons, about 60\% of the \misid is due to decays in flight, for these particular samples. The average efficiency and \misids, integrated over momentum ($p>3\,$\gevc), are also given in \tabref{tab:eimvspt}, for 5 different \pt intervals. There are not enough candidates in the muon, pion and kaon samples for a measurement dependent on momentum in the lowest \pt bin. Similarly for the protons, in the highest \pt interval.

\begin{table}
\centering\caption{Average IsMuon efficiency and \misids in different transverse momentum intervals (\%). Uncertainties are statistical.}
\label{tab:eimvspt}
\begin{tabular}{c|c|c|c|c}
$p_T$ interval (\gevc)& muon & proton & pion & kaon \\\hline
$p_T<0.8$     &                 & 1.393$\pm$ 0.005& 6.2$\pm$ 0.1 & 4.3$\pm$0.1\\ 
$0.8<p_T<1.7$ & 96.94$\pm$ 0.07 & 0.737$\pm$ 0.003& 2.19$\pm$ 0.01 & 1.93$\pm$0.1\\ 
$1.7<p_T<3.0$ & 98.53$\pm$ 0.05 & 0.149$\pm$ 0.004& 0.61$\pm$ 0.01 & 0.93$\pm$0.01\\ 
$3.0<p_T<5.0$ & 98.51$\pm$ 0.06 & 0.12$\pm$ 0.02  & 0.40$\pm$ 0.01 & 0.72$\pm$0.01\\ 
$5.0<p_T$     & 98.51$\pm$ 0.07 &                 & 0.33$\pm$ 0.02 & 0.69$\pm$0.01\\\hline 
\end{tabular}    
\end{table}

\begin{figure}[h!]
\begin{tabular}{cc}
\includegraphics[angle=0,width=0.5\textwidth]{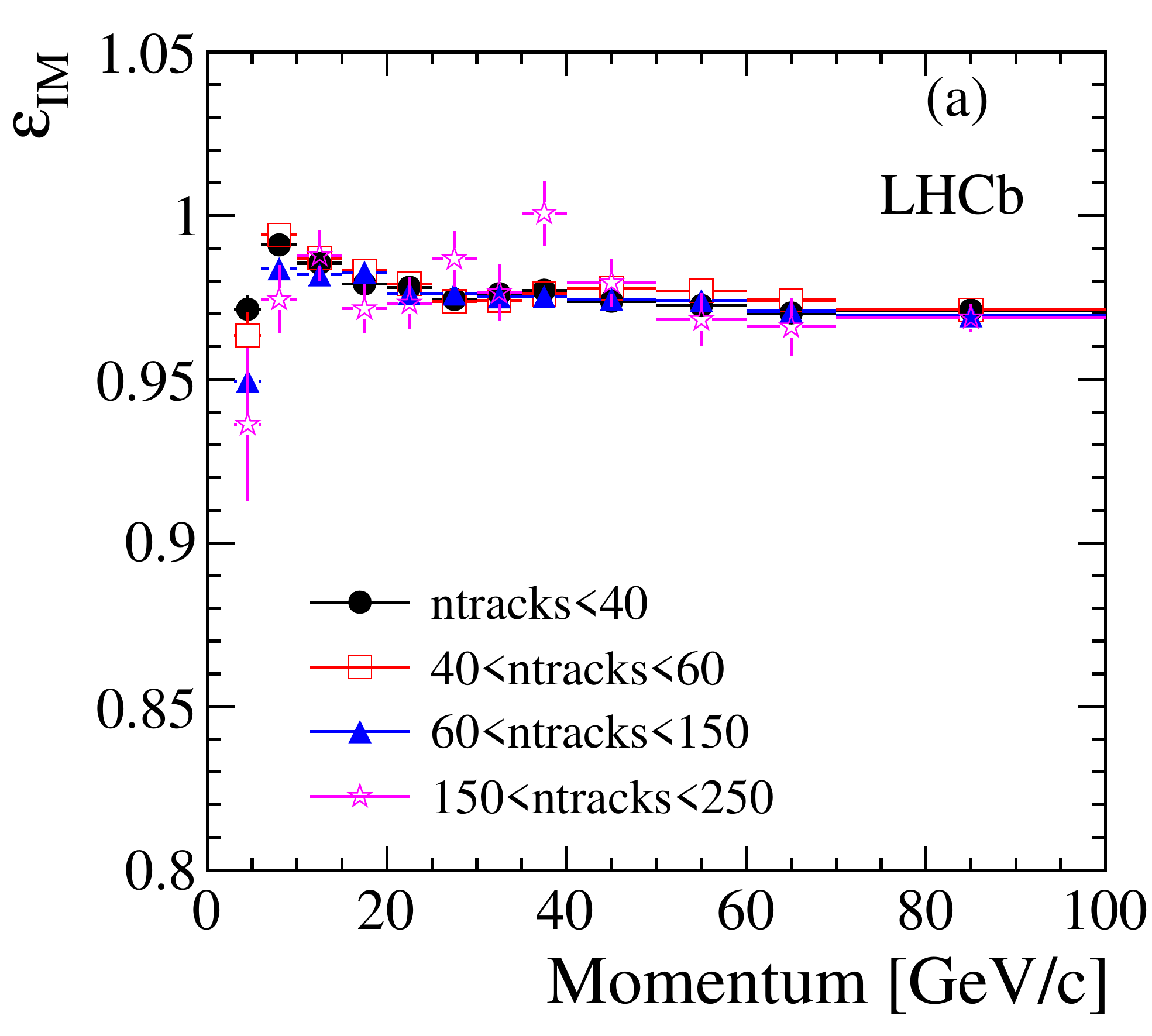} & 
\includegraphics[angle=0,width=0.5\textwidth]{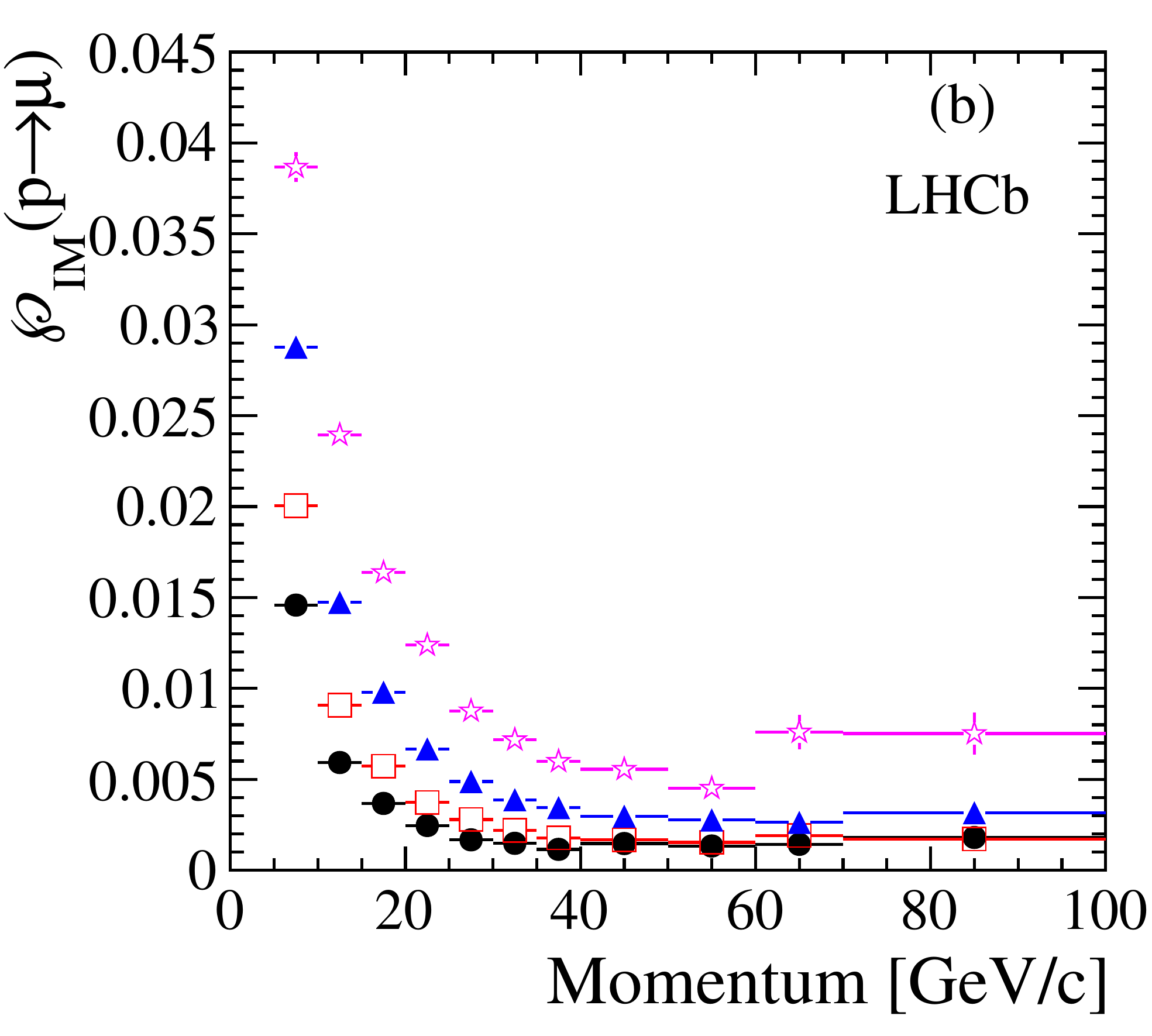} \\
\includegraphics[angle=0,width=0.5\textwidth]{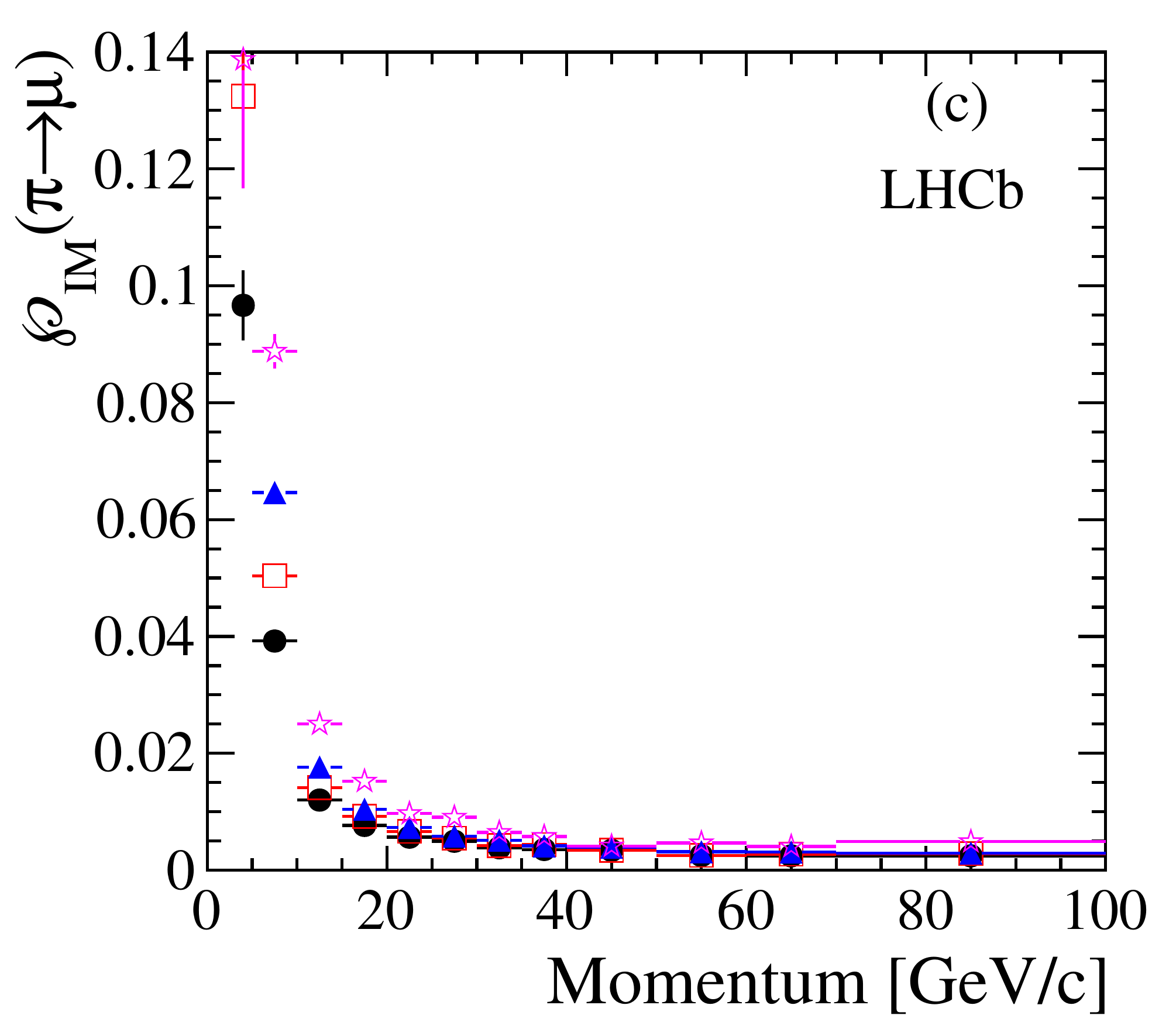} & 
\includegraphics[angle=0,width=0.5\textwidth]{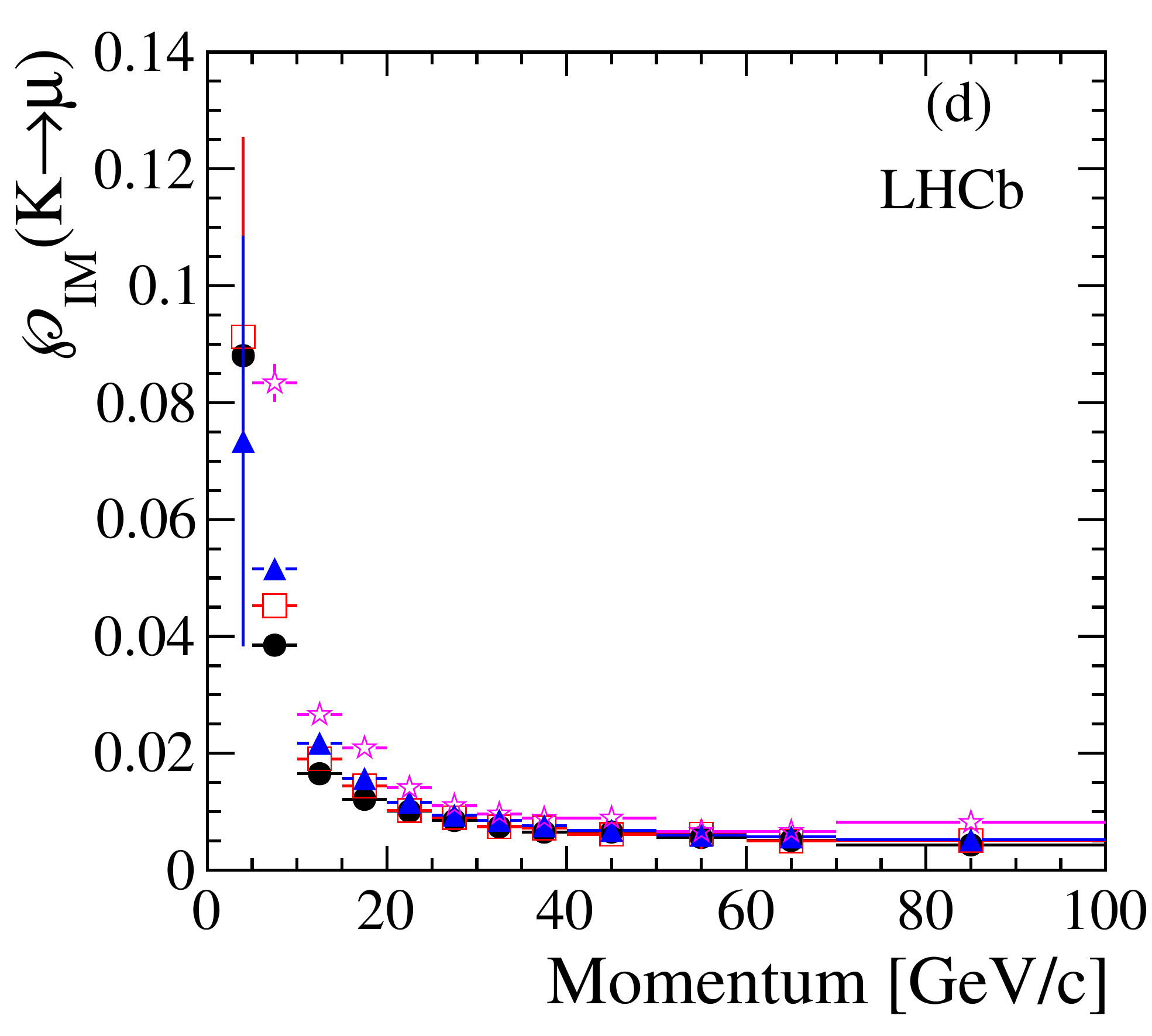} \\
\end{tabular}
  \caption{IsMuon efficiency \eIM (a) and \RIM for protons (b), pions (c) and kaons (d) as a function of momentum for different ranges of the number of trajectories reconstructed in the event (ntracks).}
  \label{fig:IMeficVSpVStrack}
\end{figure}

The LHCb detector has been designed to operate at the luminosity of ${\cal L}=2\times 10^{32}\,$cm$^{-2}$s$^{-1}$ and with a probability of having one interaction per beam crossing maximal with respect to higher numbers. However, in the 2011 run the experiment operated with an average number of interactions per beam crossing about 2.5 times the nominal average, with a corresponding increase of the overall detector occupancies. The behavior of \eIM and \RIM was then evaluated as a function of the number of tracks which contain hits in the tracking subsystems, from the VELO to the tracking stations. No significant decrease of \eIM is observed, while an increase of the \misids is seen with higher track multiplicities, as expected. The detailed behaviour of both the efficiency and the \misids as a function of momentum is shown in \figref{fig:IMeficVSpVStrack}. The probability \RIMpr increases by a factor 2.7 for particles with momentum in the range 3 to 5$\,$\gevc, when comparing events with track multiplicity smaller than 40 and events with track multiplicity between 150 and 250, which is the highest interval of multiplicity analysed. At high momentum, the difference is much less pronounced. For pions and kaons, the increase at low momentum is a factor of two, approximately,  and drops quickly to a plateau value starting at 20$\,$\gevc. Since the FOI are smaller at high momentum, the \misid becomes less sensitive to the multiplicity of the underlying event. 

The charge dependence of the efficiency \eIM is also analysed. No difference between the efficiencies is seen up to the level of the statistical fluctuations. When integrating over the whole momentum range, the relative difference is 0.09$\pm$0.08\%, compatible with zero within the statistical uncertainty.

\subsection{Performance of muon likelihoods}

\begin{figure}[htb!]
\centering \begin{tabular}{cc} 
\includegraphics[angle=0,width=0.49\textwidth]{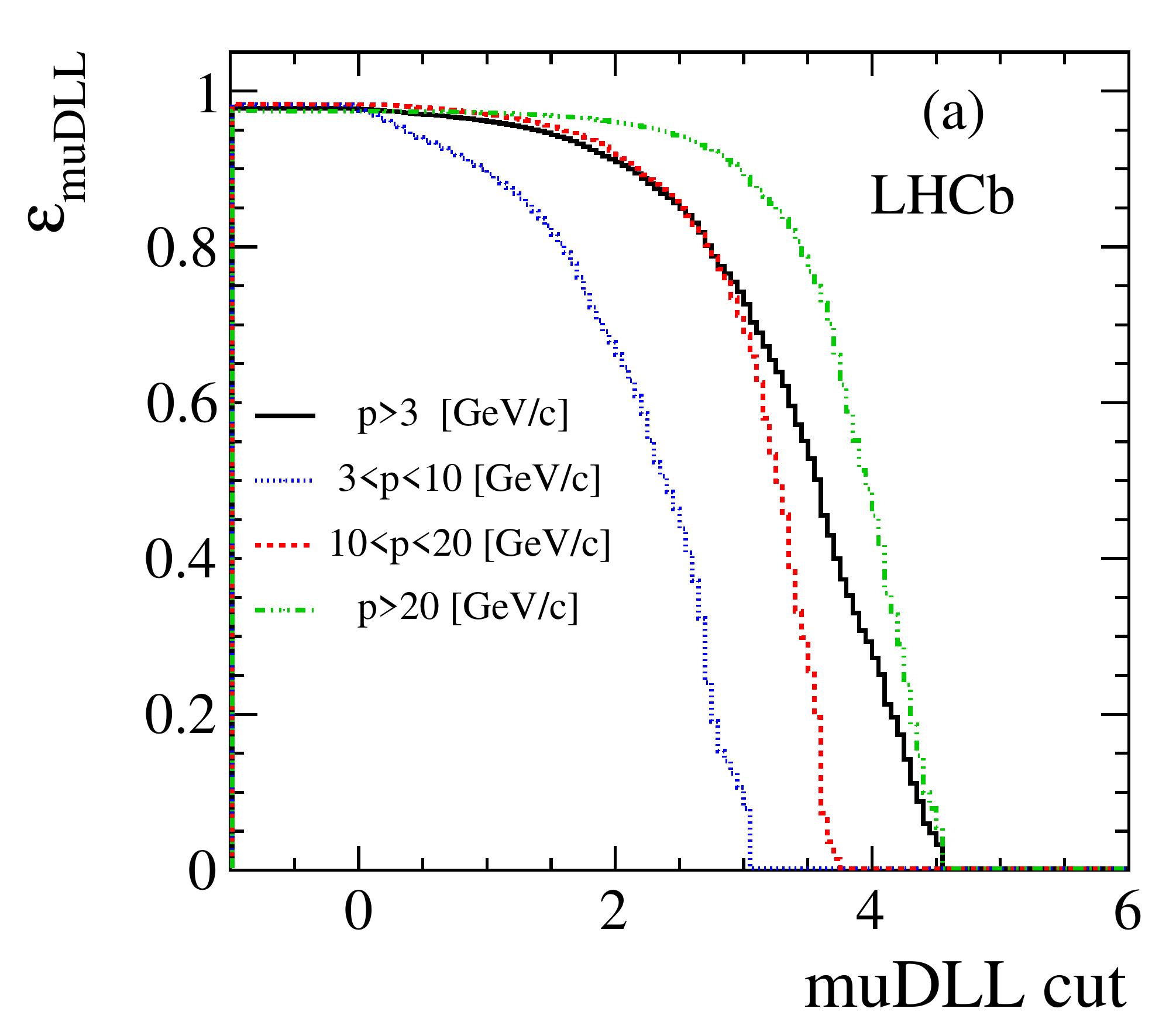} &
\includegraphics[angle=0,width=0.49\textwidth]{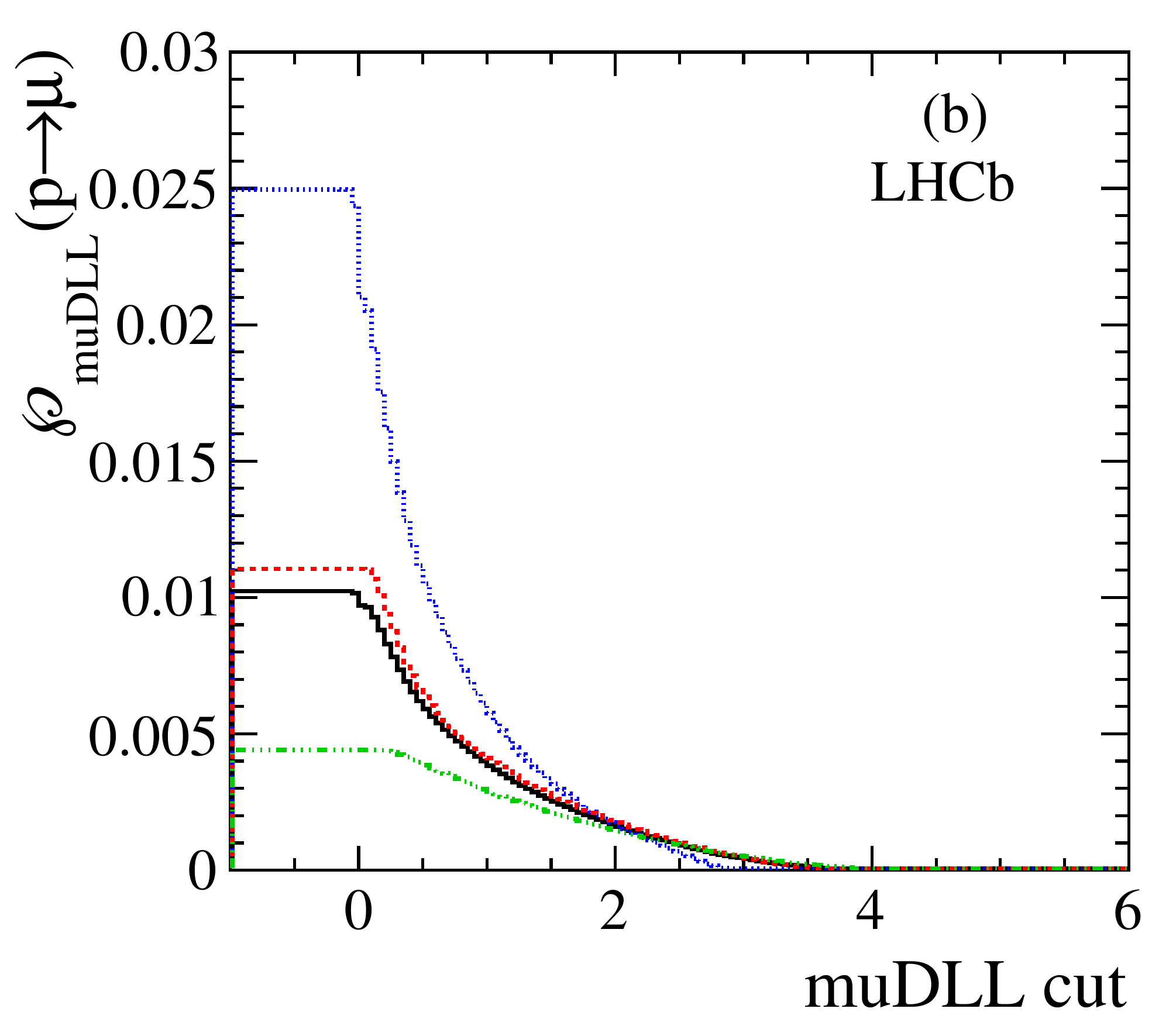}\\ 
\includegraphics[angle=0,width=0.49\textwidth]{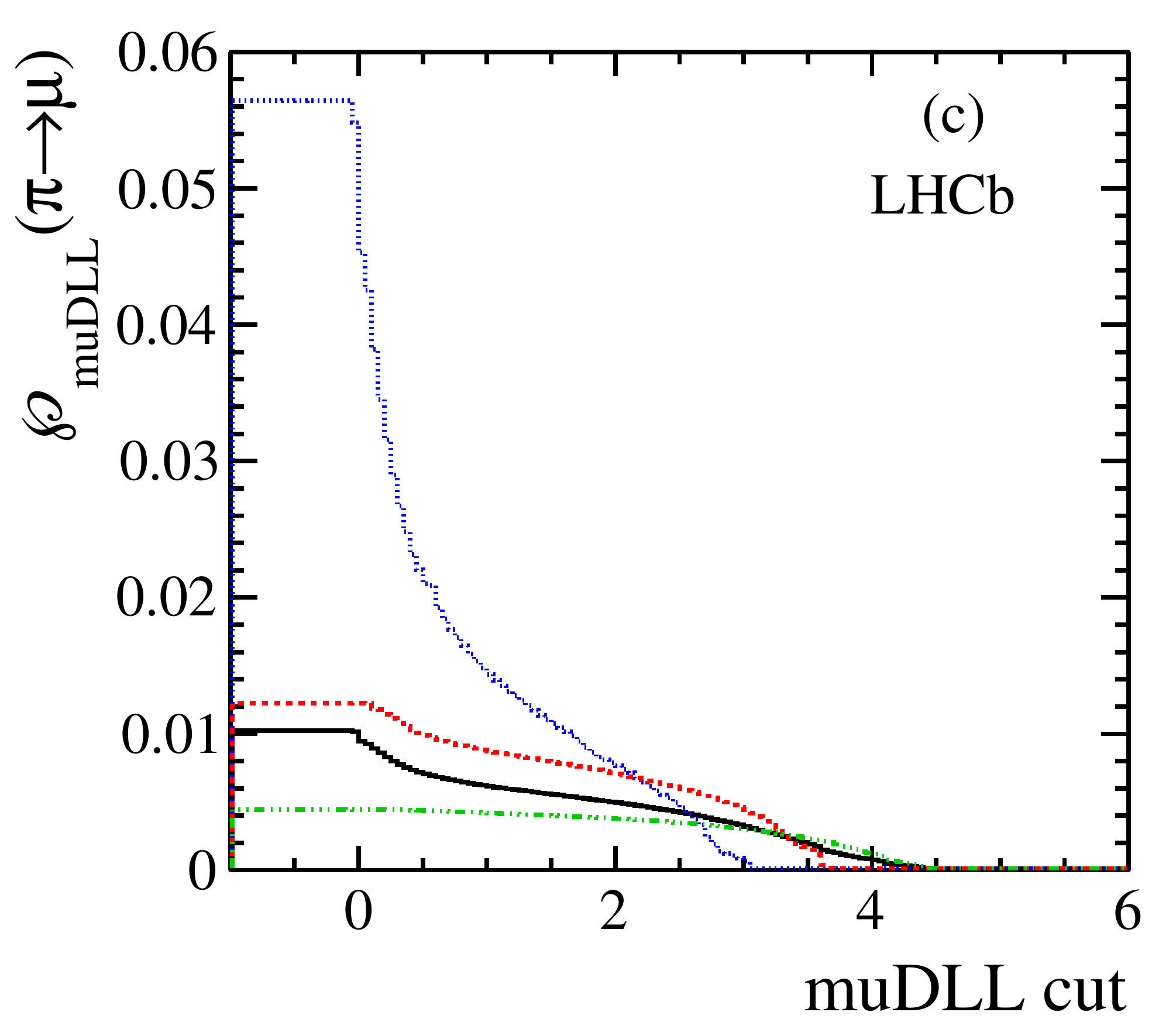}&
\includegraphics[angle=0,width=0.49\textwidth]{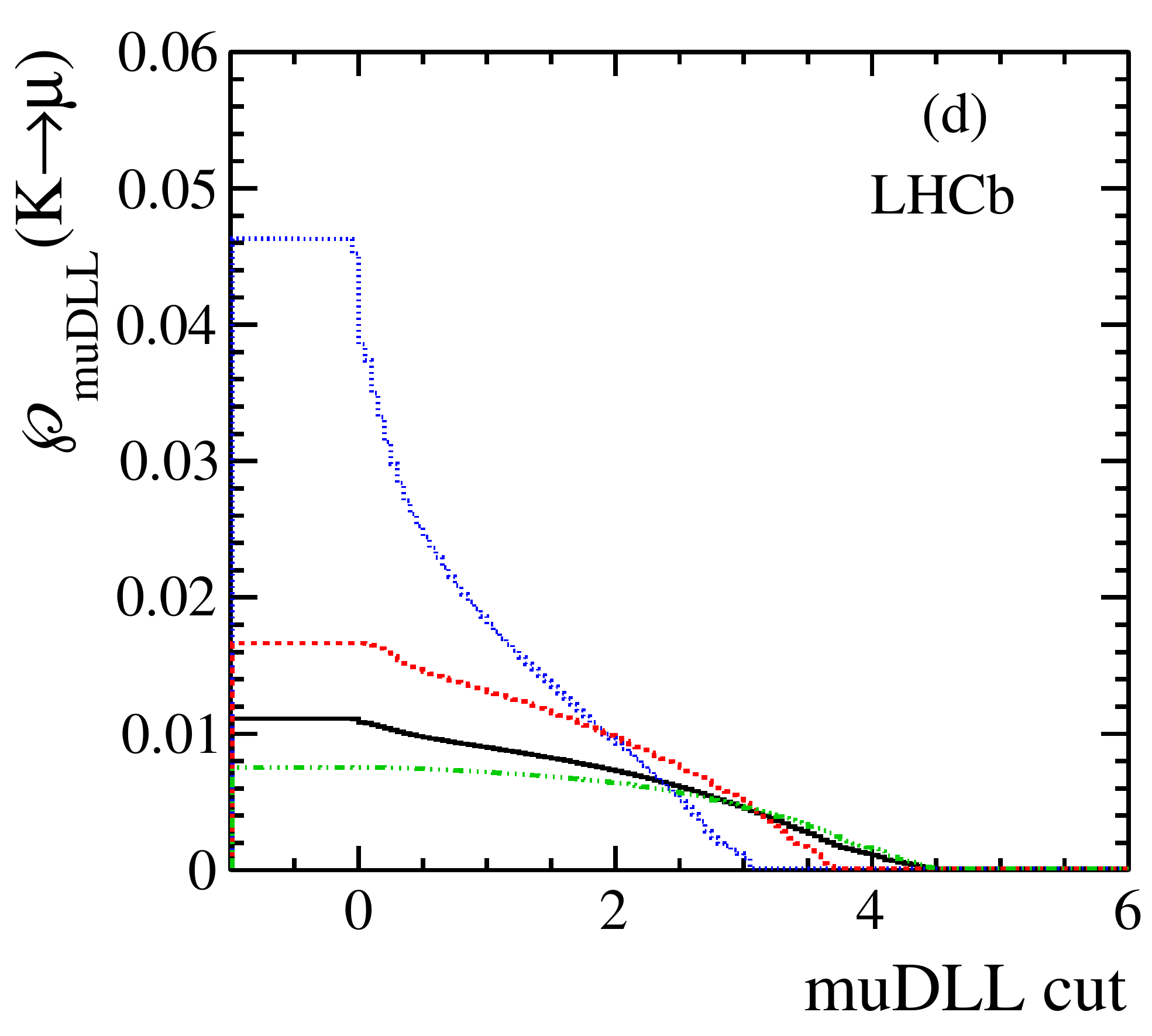}\\
\end{tabular}
\caption{The efficiency \eDLL as a function of muon DLL cut for muons (a) and \misids for protons (b), pions (c) and  kaons (d). The black solid line shows the average values integrated over $p>3\,$\gevc. The blue dotted line correspond to particles in the range $3<p<10\,$\gevc. The red dashed lines show results for $10<p<20\,$\gevc and the green dashed-dotted for $p>20\,$\gevc.}
 \label{fig:efVSDLL}
\end{figure}

The muon identification efficiency (\eDLL) is measured as a function of a selection cut in the variable muDLL, for different momentum ranges, as shown in \figref{fig:efVSDLL}(a). The \misids are also shown in \figref{fig:efVSDLL}(b) to \figref{fig:efVSDLL}(d), for the same momentum ranges. The black solid line shows the average fractions, when integrated over $p>3\,$\gevc (and \pt$>0.8\,$\gevc for the muons). All curves start at the efficiency or \misid corresponding to the IsMuon requirement. 
For tracks with $p>10\,$\gevc, the muon efficiency is independent of momentum up to muDLL$\sim$2. 
To achieve a misidentification probability independent from the momentum, the value of the muDLL cut must depend on particle momentum. By applying a muDLL cut irrespective of the momentum, the \misids show a strong momentum dependence.

As an example, when requiring muDLL$\geq$1.74, a cut that provides a final muon efficiency of 93.2\%, the final misidentification probabilities are 0.21\%, 0.78\% and 0.52\% for protons, kaons and pions respectively. This cut, which provides a sharp decrease of 5\% of the efficiency with respect to the IsMuon efficiency, is used here as an example only for a clear comparison between the muon DLL and the DLL. Since the average efficiency and \misids values are given for our calibration samples, which have their particular momentum and $p_T$ spectrum, they can be different for samples with different kinematic distributions. 

\begin{figure}[htb!]
  \centering \begin{tabular}{cc}
\includegraphics[angle=0,width=0.50\textwidth]{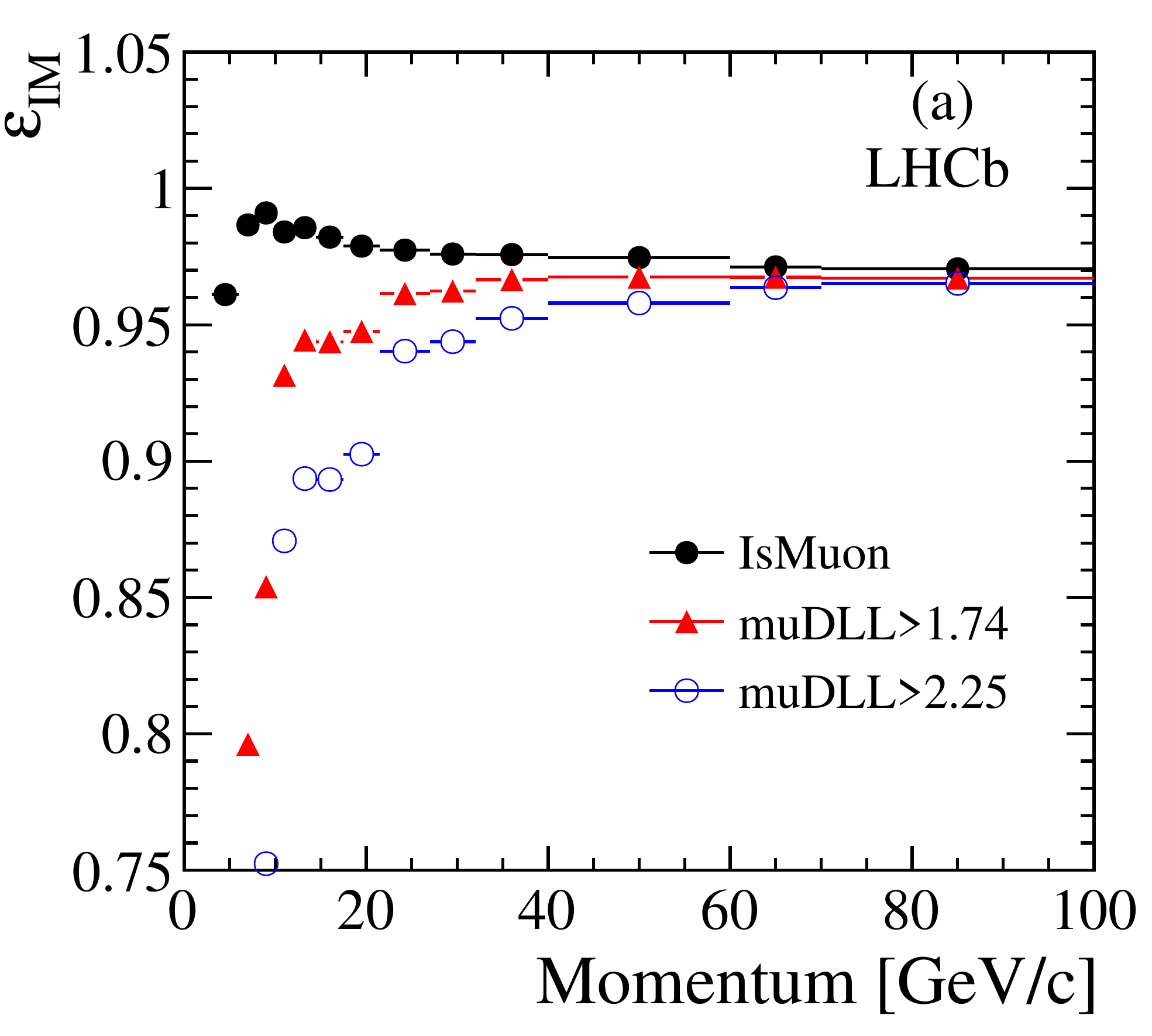}&
\includegraphics[angle=0,width=0.50\textwidth]{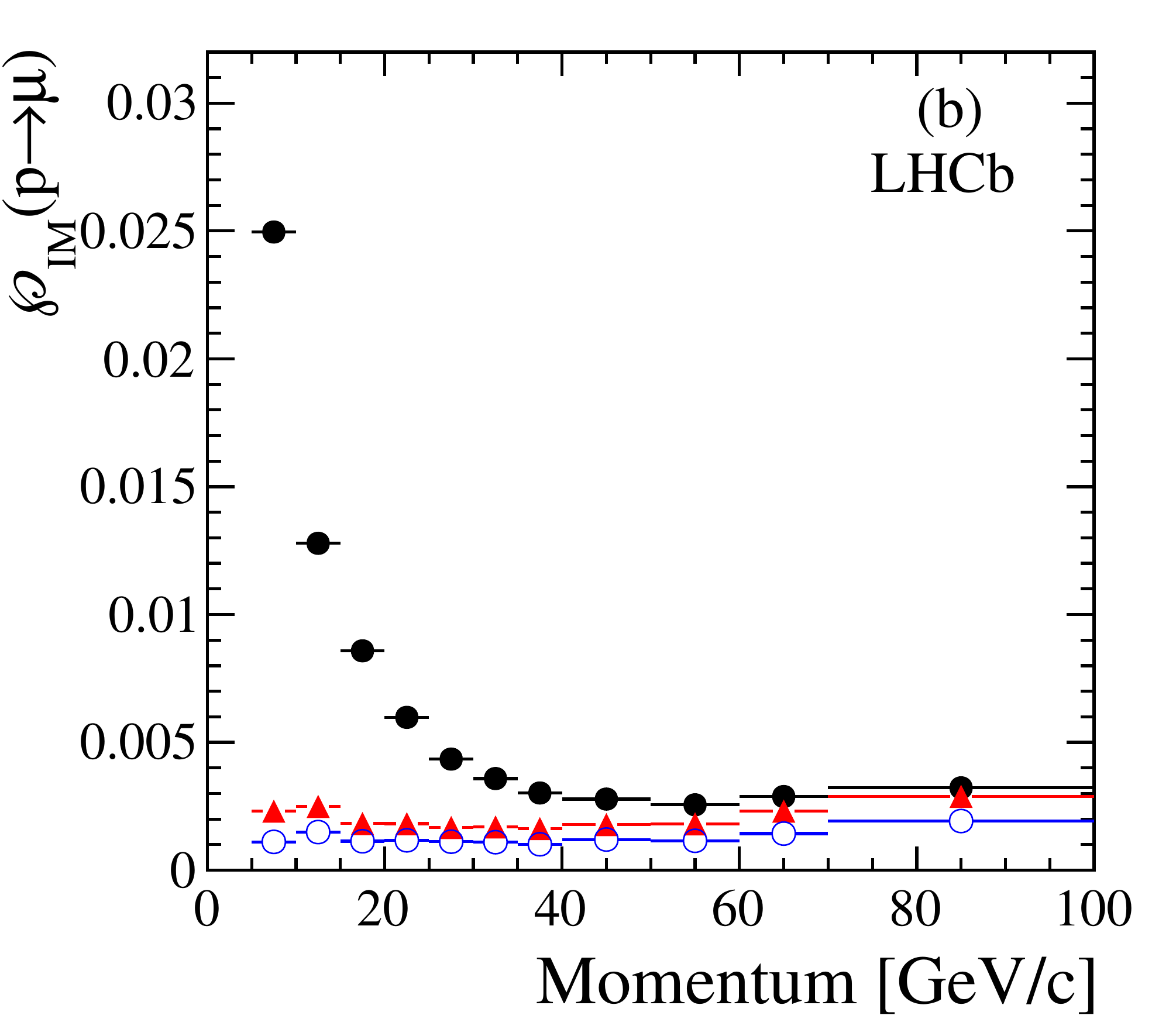}\\  
\includegraphics[angle=0,width=0.50\textwidth]{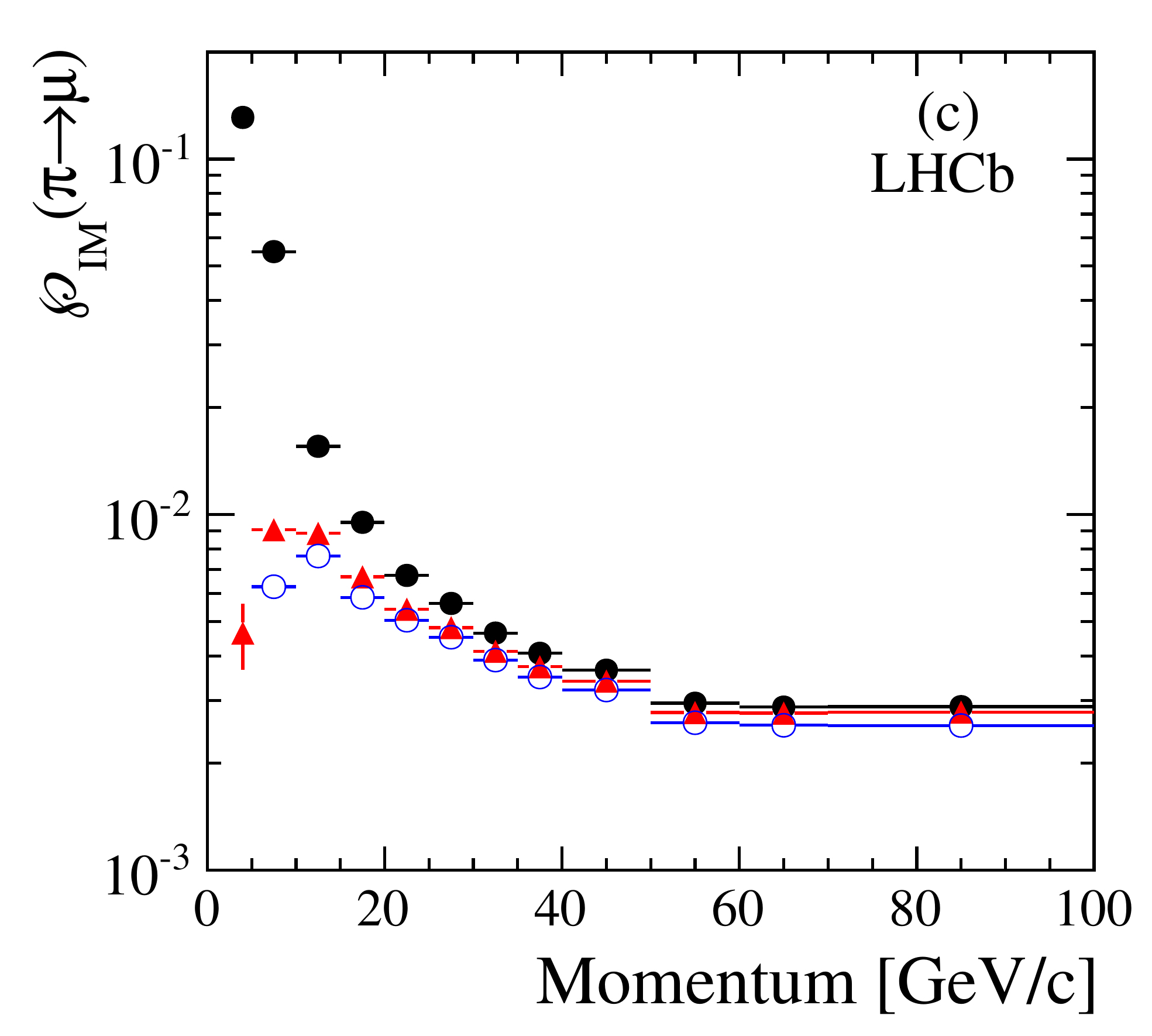}&
\includegraphics[angle=0,width=0.50\textwidth]{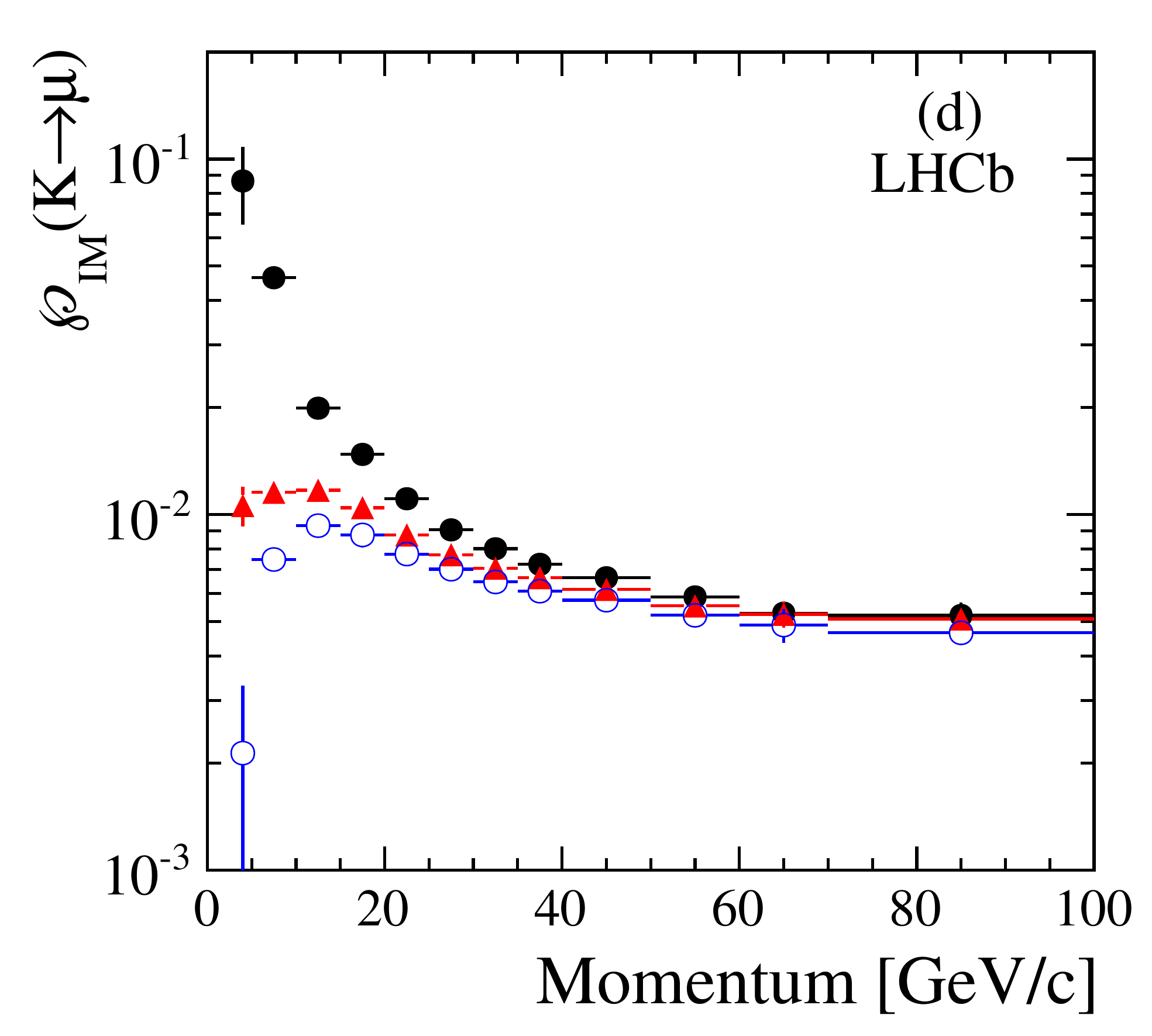}\\  
\end{tabular}
  \caption{Muon efficiency (a) and \misids for protons (b), pions (c) and kaons (d) as a function of the particle momentum for the IsMuon requirement alone (black solid circles) and with the additional cuts muDLL$\geq$1.74 (red triangles) and muDLL$\geq$2.25 (blue open circles).}  
 \label{fig:emudllvsp}
\end{figure}

The momentum dependence of \eDLL and of \RDLL for particles satisfying this particular cut, muDLL$\geq$1.74, are shown in \figref{fig:emudllvsp}, compared to the IsMuon requirement alone and a tighter selection, muDLL$\geq$2.25. Again, this second cut was chosen for providing a sharp reduction of the muon efficiency of 10\% with respect to the IsMuon efficiency. Once more, since the performance is integrated over $p_T$, small variations from these values are expected for different samples, in particular for the \misids, which present a stronger dependence with transverse momentum.

\subsection{Performance of combined likelihoods}

\begin{figure}[htb!]
  \centering \begin{tabular}{cc}
\includegraphics[angle=0,width=0.45\textwidth]{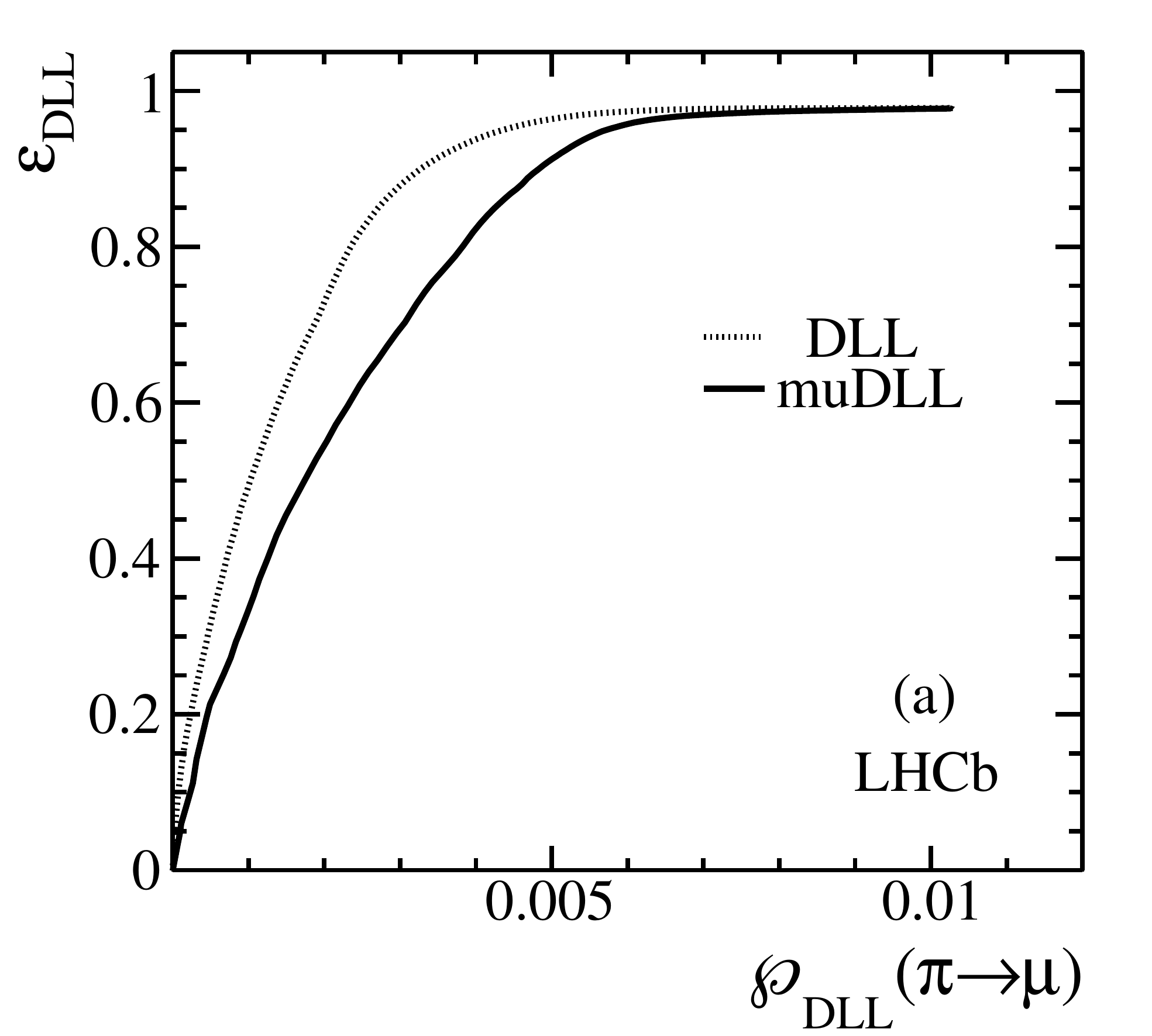}&
\includegraphics[angle=0,width=0.45\textwidth]{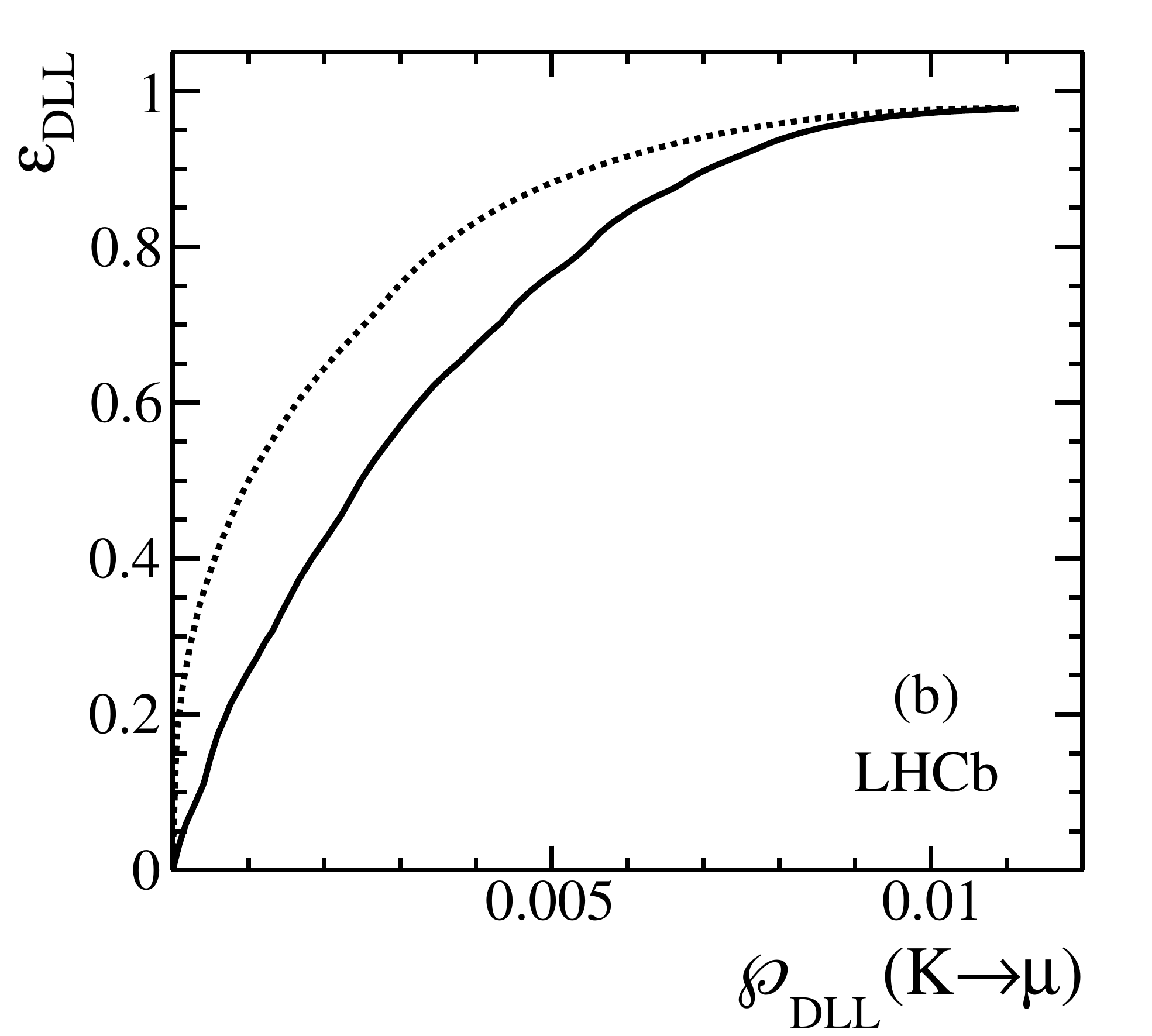}\\
\end{tabular}
  \caption{Average efficiency \eCDLL as a function of the pion (a) and  kaon (b) \misids for particles with momentum in the range $p>3$ $\,$\gevc. The dotted lines show the DLL performance, while the  muon DLL performance is shown with a solid line.} 
 \label{fig:cdll_roc}
\end{figure}

The DLL efficiency is shown as a function of the pion and kaon \misids in \figref{fig:cdll_roc}, together with the results obtained using the muDLL alone, allowing for a direct comparison of their performances. 

The DLL benefits from RICH and calorimeter information, being more effective than the muon DLL alone in separating pions and kaons from muons. After IsMuon, this is the most used particle identification requirement used to select muons in LHCb and the actual cut value is usually chosen according to the compromise between purity and efficiency needed for that specific study. 
The average misidentification rates corresponding to a cut which provides an average decrease of 5\% (equivalent to the one obtained with muDLL$\geq$1.74, as previously shown) are around 0.65\% and 0.38\% for the kaons and pions, respectively.

\subsection{Performance of selections based on hits sharing}

As mentioned in \secref{sec:principles}, after requiring IsMuon, an additional way of reducing the incorrect identification probability of hadrons as muons, in particular at high occupancy, is the use of a cut on NShared. 
 
\begin{figure}
\centering{
\begin{tabular}{cc}
\includegraphics[angle=0,width=0.45\textwidth]{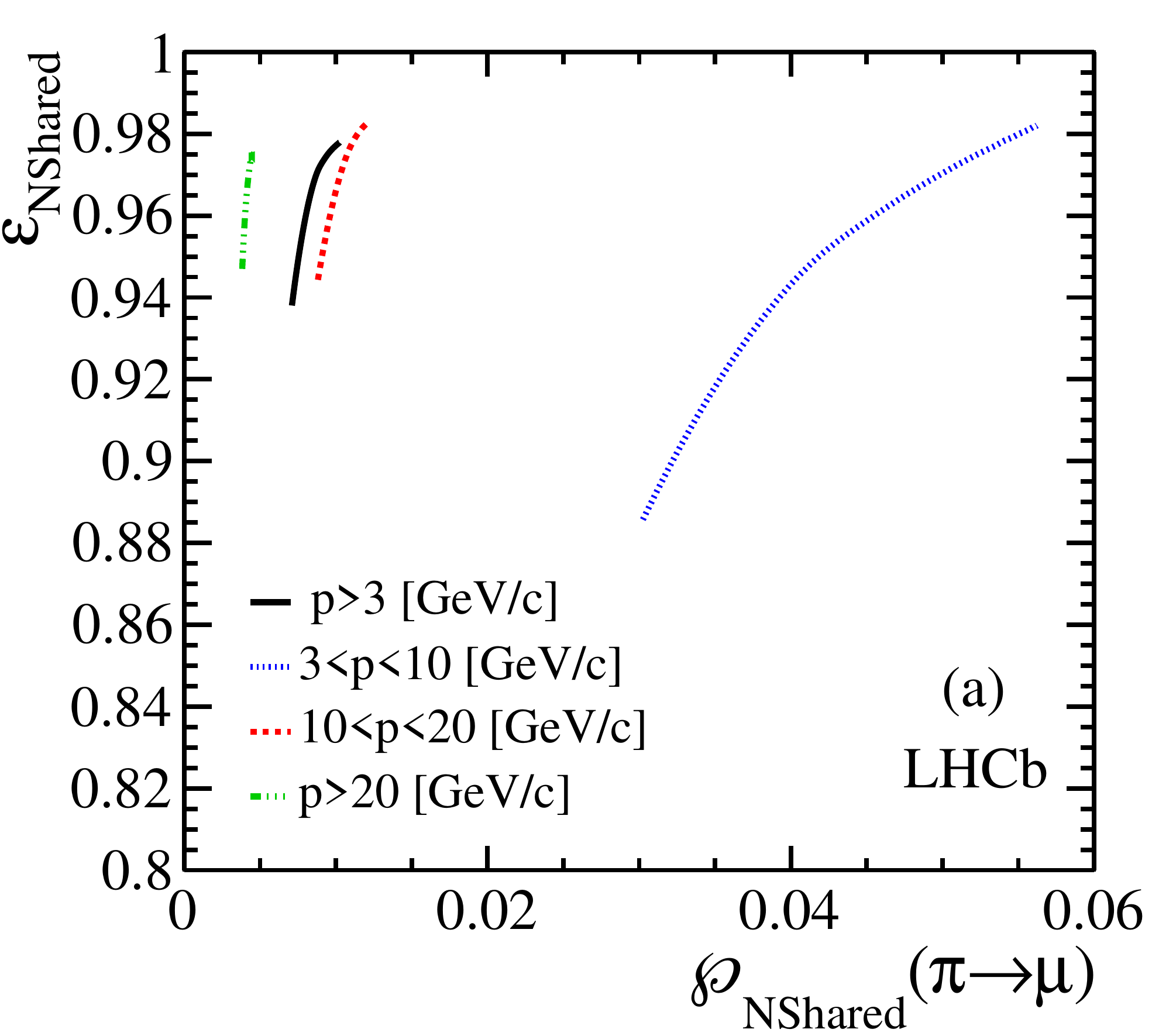}& 
\includegraphics[angle=0,width=0.45\textwidth]{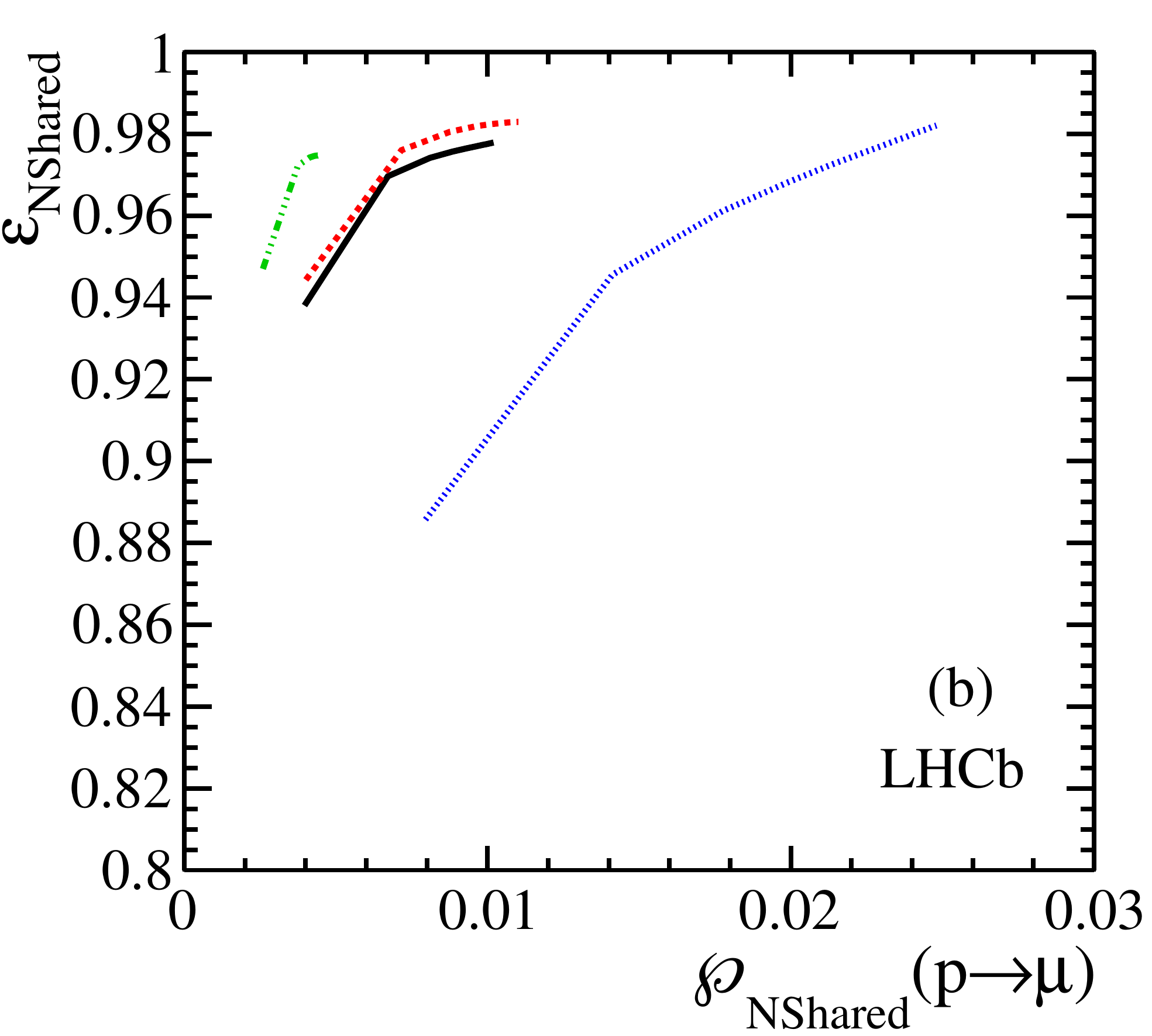}\\
\end{tabular}}
\caption{Muon efficiency \ensh as a function of the pion and proton \misids. The average values, for all particles with $p>3\,$\gevc, are shown with a black line, compared to the three momentum ranges separately, as for 
Fig.~\protect\ref{fig:efVSDLL}.} 
\protect\label{fig:nsh_roc}
\end{figure}

The muon efficiency is shown as a function of the pion \misid for corresponding NShared cut in \figref{fig:nsh_roc}(a); protons are shown in \figref{fig:nsh_roc}(b). Due to similar decay-in-flight pollution at low momentum, kaons behave as pions.  The black solid line shows the average values integrated over $p>3\,$\gevc. The NShared selection is particularly effective at low momenta, with increasing the FOI size.

\input{systematics}

%% file: systematics.tex
\subsection{Systematic checks}
\label{sec:systematics}

The effect of the trigger and of the method chosen to evaluate the efficiency and misidentificatin probabilities are investigated. 
Alternatively to the requirement of the \Jpsimumu sample being triggered independently of the probe muon, a muon trigger decision based on the tag muon was used to evaluate the IsMuon efficiency. The systematic uncertainty due to the choice of trigger strategy is taken as the difference between the two determinations, which is 0.2\%.

When performing a full fit to the signal and background components of the mass distributions used to extract the yields of signal events satisfying or not the muon identification requirements, the resulting efficiencies and proton \misid rates agree within the statistical uncertainties with the results shown in \secref{sec:results}. 
  
For the pion and kaon misidentification probabilities, the effect of the trigger is studied and found to be negligible within the uncertainties, independently of momentum and transverse momentum.
Also the systematic uncertainty related to the method used for the evaluation of the efficiency is found to be negligible as a function of momentum, apart from a few intervals where it is comparable with the statistical accuracy.

%% file: conclusion.tex
\section{Conclusions}
\label{sec:conclusions}

The performance of the muon identification procedure used in the LHCb experiment has been evaluated, using a dataset corresponding to 1\invfb recorded in 2011 at $\sqrt{s}=7\,$\tev. 

A loose binary criterium that can be used to select muons is based on the matching of muon hits with the particle trajectory. For candidates satisfying this requirement, likelihoods for muon and non-muon hypotheses are built with the pattern of hits around the trajectories, which can be used to refine the selection. An additional way of rejecting fake muon candidates is provided by a variable sensitive to hit sharing by nearby particles. 

The muon identification efficiency was observed to be robust against the variation of detector occupancies and presents a weak dependence on momentum and transverse momentum. Hadron \misids present a stronger dependence on hit or track multiplicity, however the highest increase factors are observed only for low momentum particles. 

Average muon identification efficiencies at the 98\% level are attainable for pion and kaon misidentification below the 1\% level at high transverse momentum, using the loosest identification criterium. The performance of additional requirements based on likelihoods or on hits sharing can be tuned according to the needs of each analysis and reduce the \misids dependence on track multiplicity. Adding a requirement on the difference of the log-likelihoods that provides a total muon efficiency at the level of 93\%, the hadron \misids are below 0.6\%.